\documentclass[twocolumn]{svjour3}

\smartqed

\usepackage[T1]{fontenc}
\usepackage[utf8]{inputenc}

\usepackage{algorithm}
\usepackage{algpseudocode}
\usepackage{amsmath}
\usepackage{amssymb}
\usepackage{booktabs}
\usepackage{colortbl}
\usepackage{graphicx}
\usepackage{hyperref}
\usepackage{mathtools}
\usepackage{multirow}
\usepackage{pifont}
\usepackage[caption=false]{subfig}
\usepackage{tcolorbox}
\usepackage{threeparttable}
\usepackage{xcolor}

\renewcommand{\vec}[1]{\mathbf{#1}}

\DeclarePairedDelimiter{\norm}{\lVert}{\rVert}

\newcommand{\KNN}{$k$-NN}
\newcommand{\ANN}{ANN}
\newcommand{\KDTREE}{$k$\nobreakdash-d~tree}

\newcommand{\cmark}{\ding{51}}
\newcommand{\xmark}{\ding{55}}

\newcolumntype{a}{>{\columncolor[gray]{0.9}}c}
\newcommand{\ch}{High}
\newcommand{\cm}{Med.}
\newcommand{\cl}{Low}

\newcommand{\hi}[1]{\vspace{.25em}\noindent {\textbf{#1}}}

\begin{document}

\title{Survey of Vector Database Management Systems}

\author{James Jie Pan \and Jianguo Wang \and Guoliang Li}

\institute{
James Jie Pan \at
Department of Computer Science and Technology \\
Tsinghua University, Beijing, China \\
\email{jamesjpan@tsinghua.edu.cn}
\and
Jianguo Wang \at
Department of Computer Science \\
Purdue University, West Lafayette, Indiana, USA \\
\email{csjgwang@purdue.edu}
\and
Guoliang Li \at
Department of Computer Science and Technology \\
Tsinghua University, Beijing, China \\
\email{liguoliang@tsinghua.edu.cn}
}

\date{Received: date / Accepted: date}

\maketitle

\setcounter{tocdepth}{3}  

\begin{abstract}
There are now over 20 commercial vector database management systems
(VDBMSs), all produced within the past five years. But embedding-based
retrieval (EBR) has been studied for over ten years, and similarity
search a staggering half century and more.
Driving this shift from algorithms to systems are new data intensive
applications, notably large language models (LLMs), that demand vast
stores of unstructured data coupled with reliable, secure, fast, and
scalable query processing capability.
A variety of new data management techniques now exist for addressing
these needs, however there is no comprehensive survey to thoroughly
review these techniques and systems.

We start by identifying five main obstacles to vector data management,
namely the vagueness of semantic similarity, large size of vectors,
high cost of similarity comparison, lack of a natural partitioning
that can be used for indexing, and difficulty of efficiently answering
``hybrid'' queries that require both attributes and vectors.
Overcoming these obstacles has led to new approaches to query
processing, storage and indexing, and query optimization and
execution.
For query processing, a variety of similarity scores and query types
are now well understood; for storage and indexing, techniques include
vector compression, namely quantization, and partitioning techniques
based on randomization, learning partitioning, and ``navigable''
partitioning; for query optimization and execution, we describe new
operators for hybrid queries, as well as techniques for plan
enumeration, plan selection, and hardware accelerated query execution.
These techniques lead to a variety of VDBMSs across a spectrum of
design and runtime characteristics, including ``native'' systems that
are specialized for vectors and ``extended'' systems that incorporate
vector capabilities into existing systems.
We then discuss benchmarks, and finally we outline several research
challenges and point the direction for future work.
\end{abstract}


\section{Introduction}
\label{sec:intro}

\begin{figure}[t]
\centering
\includegraphics[width=0.40\textwidth]{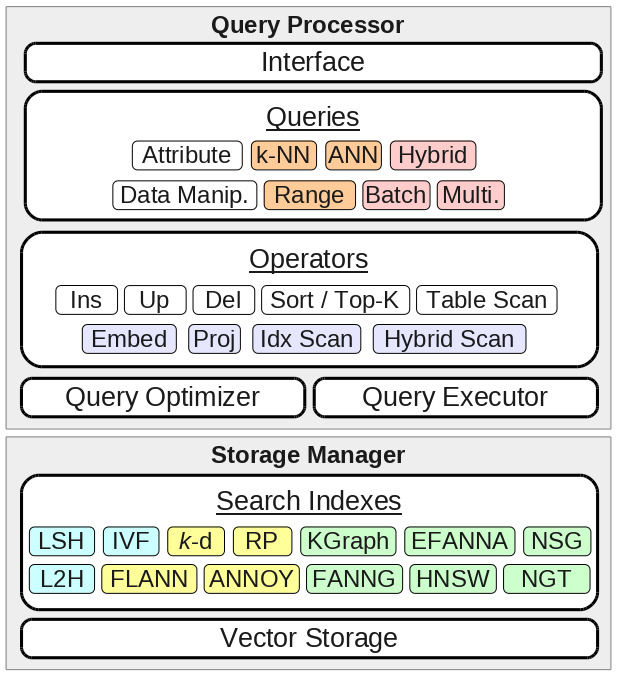}
\caption{Architecture of a VDBMS.}
\label{fig:arch}
\end{figure}

The rise of large language models (LLMs) \cite{jurafsky2009} for tasks
like information retrieval \cite{asai2023}, along with the growth of
unstructured data underlying economic drivers such as e-commerce and
recommendation platforms \cite{wei2020,wang2021,guo2022}, calls for
new \emph{vector database management systems} (VDBMSs) that can
deliver traditional capabilities such as query optimization,
transactions, scalability, fault tolerance, and privacy and security,
but for unstructured data.

As these data are not represented by attributes from a fixed schema,
they are not retrieved through structured queries but through
similarity search, where data that have similar semantic meanings to
the query are retrieved \cite{mitra2018}. To support this type of
search, entities such as images and documents are first encoded into
$D$-dimensional feature vectors via an embedding model before being
stored inside a VDBMS. The dual-encoder model \cite{chang2020}
describes this procedure, also known as dense retrieval
\cite{kim2022}.

Consequently, the modules in a VDBMS split into a \emph{query
processor}, which includes the query specifications, logical
operators, their physical implementations, and the query optimizer;
and the \emph{storage manager}, which maintains the search indexes and
manages the physical storage of the vectors. This is illustrated
in Figure~\ref{fig:arch}.

The designs of these modules affect the runtime characteristics of the
VDBMS. Many applications such as LLMs are read-heavy, requiring
high query throughput and low latency.
Others such as e-commerce are also write-heavy, requiring high
write throughput. Additionally, some applications require high query
accuracy, meaning that retrieved entities are true semantic matches to
the query, while other applications may be more tolerant of
errors. Developing a suitable VDBMS therefore requires understanding
the landscape of techniques and how they affect characteristics of the
system.

While there is mature understanding of processing conventional
structured data, this is not the case for vector data. We present
five key obstacles.
(1) \emph{Vague Search Criteria.} Structured queries use precise
boolean predicates, but vector queries rely on a vague notion of
semantic similarity that is hard to accurately capture.
(2) \emph{Expensive Comparisons.} Attribute predicates (\textit{e.g.} $<$,
$>$, $=$, and $\in$) can mostly be evaluated in $O(1)$ time, but a
similarity comparison typically requires $O(D)$ time, where $D$ is the
vector dimensionality.
(3) \emph{Large Size.} A structured query usually only accesses a
small number of attributes, making it possible to design
read-efficient storage structures such as column stores. But vector
search requires full feature vectors. Vectors sometimes even span
multiple data pages, making disk retrievals more expensive while also
straining memory.
(4) \emph{Lack of Structure.} Structured attributes are mainly
sortable or ordinal, leading to partitionings via numerical ranges or
categories that can be used to design search indexes. But vectors have
no obvious sort order nor are they ordinal, making it hard to design
indexes that are both accurate and efficient.
(5) \emph{Incompatibility with Attributes.}  Structured queries over
multiple attribute indexes can use simple set operations, such as
union or intersection, to collect the intermediate results into the
final result set. But vector indexes typically stop after finding $k$
most similar vectors, and combining these with the results from an
attribute index scan can lead to fewer than expected results.  On the
other hand, modifying the index scan operator to account for attribute
predicates can degrade index performance. It remains unclear how to
support ``hybrid'' queries over both attributes and vectors in a way
that is both efficient and accurate.

There are now a variety of techniques that have been developed around
these issues, aimed at achieving low query latency, high result
quality, and high throughput while supporting large numbers of
vectors. Some of these are results of decades of study on similarity
search. Others, including hybrid query processing, indexes based on
vector compression, techniques based on hardware acceleration, and
distributed architectures are more recent inventions.

In this paper, we start by surveying these techniques from the
perspective of a generic VDBMS, dividing them into those that apply to
query processing and those that apply to storage and indexing. Query
optimization and execution are treated separately from the core query
processor. Following these discussions, we apply our understanding of
these techniques to characterize existing VDBMSs.

\hi{Query Processing.}
The query processor mainly deals with how to specify the search
criteria in the first place and how to execute search queries.
For the former, a variety of similarity scores, query types, and query
interfaces are available.
For the latter, the basic operator is similarity projection, but as
it can be inefficient, a variety of index-supported operators have
been developed.
We discuss the query processor in Section~\ref{sec:proc}.

\hi{Storage and Indexing.}
The storage manager mainly deals with how to organize and store the
vector collection to support efficient and accurate search. For most
systems, this is achieved through vector search indexes.
We classify indexes into \emph{table-based} indexes such as E$^2$LSH
\cite{datar2004}, SPANN \cite{chen2021}, and IVFADC
\cite{jegou2011product}, that are generally easy to update;
\emph{tree-based} indexes such as FLANN \cite{muja2009}, RPTree
\cite{dasgupta2008,dasgupta2013}, and ANNOY \cite{annoy} that aim to
provide logarithmic search; and \emph{graph-based} indexes such as
KGraph \cite{dong2011}, FANNG \cite{harwood2016}, and HNSW
\cite{malkov2020} that have been shown to perform empirically well
but with less theoretical understanding.

To address the difficulty of partitioning vector collections,
techniques include randomization
\cite{indyk1998,datar2004,andoni2015,muja2009,dasgupta2013,dong2011,%
wang2012,subramanya2019}, learned partitioning
\cite{wang2018,jegou2011product,matsui2018,muja2009,silpa2008}, and
what we refer to as ``navigable'' partitioning
\cite{dearholt1988,malkov2014,malkov2020}.
To deal with large storage size, several techniques have been
developed for indexes over compressed vectors, including quantization
\cite{gray1984,jegou2011product,matsui2018,sivic2003,wang2020,wei2020},
as well as disk-resident indexes
\cite{gollapudi2023,chen2021}. We discuss indexing in
Section~\ref{sec:indexing}.

\hi{Optimization and Execution.}
The query optimizer and executor mainly deals with plan enumeration,
plan selection, and physical execution. To support hybrid queries,
several hybrid operators have been developed, based on what we refer
to as ``block-first'' scan \cite{wei2020,wang2021,gollapudi2023} and
``visit-first'' scan \cite{wu2022}. There are also several techniques
for enumeration and selection, including rule-based and cost-based
selection \cite{wei2020,wang2021}. For query execution, several
techniques aim to exploit the storage locality of large vectors to
design hardware accelerated operators, taking advantage of
capabilities such as processor caches \cite{wang2021}, SIMD
\cite{wang2021,andre2017,andre2021}, and GPUs
\cite{johnson2021}. There are also distributed search techniques and
techniques for supporting high throughput updates, namely based on
out-of-place updates. We discuss optimization and execution in
Section~\ref{sec:opt}.

\hi{Current Systems.}
We classify existing VDBMSs into \emph{native} systems which are
designed specifically around vector management, including Vearch
\cite{li2018}, Milvus \cite{wang2021}, and Manu \cite{guo2022};
\emph{extended} systems which add vector capabilities on top of an
existing data management system, including AnalyticDB-V \cite{wei2020}
and PASE \cite{yang2020}; and \emph{search engines and libraries}
which aim to provide search capability only, such as Apache Lucene
\cite{lucene}, Elasticsearch \cite{elastic}, and Meta Faiss
\cite{faiss}.
Native systems tend to favor high performance techniques targeted at
specific capabilities, while extended systems tend to favor techniques
that are more adaptable to different workloads but are not necessarily
the fastest.
We survey current systems in Section~\ref{sec:sys}.

\hi{Related Surveys.}
A high level survey is
available\footnote{\url{http://arxiv.org/abs/2309.11322}} that mostly
focuses on fundamental VDBMS concepts and use cases. Likewise, some
tutorials are available that focus specifically on similarity search
\cite{qin2020,qin2021}. We complement these by focusing on specific
problems and techniques related to vector data management as a whole.
Surveys are also available covering data types that are related to
vectors, such as time series and strings, but not supported by VDBMSs.
Unlike systems for these other data types, a VDBMS can make no
assumptions\footnote{\textit{E.g.} the correlations in a time series.}
about feature vector dimensions. We refer readers to
\cite{echihabi2019,echihabi2021}.

For the remaining sections, we briefly discuss benchmarks in
Section~\ref{sec:bench}, followed by a summary of research challenges
and open problems in Section~\ref{sec:disc}. We conclude the survey in
Section~\ref{sec:conc}.


\section{Query Processing}
\label{sec:proc}

Query processing in a VDBMS begins with a search specification
consisting of the \emph{similarity score} and the \emph{query
type}. Search criteria are conveyed to the system through a
\emph{query interface}. Once a query is received by the system, it is
processed by executing a chain of operators over the vector
collection.

\subsection{Similarity Scores}

A similarity score
$f:\mathbb{R}^D\times\mathbb{R}^D\rightarrow\mathbb{R}$ maps two
$D$-dimensional vectors, $\vec{a}$ and $\vec{b}$, onto a scalar,
$f(\vec{a},\vec{b})$, with larger values indicating greater
similarity.

Similarity is often measured via distance in practice, with values
closer to $0$ indicating greater similarity. Distance functions obey
the metric axioms of identity ($d(\vec{a},\vec{a})=0$), positivity
($d(\vec{a},\vec{b})>0\textrm{ if }\vec{a}\neq\vec{b}$), symmetry
($d(\vec{a},\vec{b})=d(\vec{b},\vec{a})$), and triangle inequality
($d(\vec{a},\vec{c})\leq d(\vec{a},\vec{b})+d(\vec{b},\vec{c})$ for
any three vectors $\vec{a},\vec{b},\vec{c}$).
Several similarity scores are commonly supported by VDBMSs, and these
are summarized in Table~\ref{tab:scores}.

Aside from these basic scores, some VDBMSs also support
\emph{aggregate scores} for applications like multi-vector search
\cite{wang2020}. There is also emerging work on \emph{learned scores}
\cite{abdelkader2019,zhang2020,meng2022},
but these are not available in commercial systems.

\begin{table}[t]
\centering
\caption{Common similarity scores.}
\label{tab:scores}
\begin{tabular}{lllll}
\toprule
Type
  & Score
  & Metric
  & Complexity
  & Range \\
\midrule
Sim.
  & Inner Prod.
  & \xmark
  & $O(D)$
  & $\mathbb{R}$ \\

  & Cosine
  & \xmark
  & $O(D)$
  & $\left[-1,1\right]$ \\
Dist.
  & Minkowski
  & \cmark
  & $O(D)$
  & $\mathbb{R}^+$ \\

  & Mahalanobis
  & \cmark
  & $O(D^{2+O(1)})$
  & $\mathbb{R}^+$ \\

  & Hamming
  & \cmark
  & $O(D)$
  & $\mathbb{N}$ \\
\bottomrule
\end{tabular}
\end{table}

\subsubsection{Basic Scores}

The Hamming distance counts the number of differing dimensions between
vectors $\vec{a}$ and $\vec{b}$.

\begin{definition}[Hamming]
$d(\vec{a},\vec{b})=\sum_{i=1}^n \delta_{a_ib_i}$
\end{definition}
where $\delta$ is the Kronecker delta.

%
Other similarity scores start from an inner product.
For $\mathbb{R}^D$ space, the dot product typically serves as the
inner product.

\begin{definition}[Inner Product]
\label{def:dot}
$f(\vec{a},\vec{b})=\sum_{i=1}^n a_ib_i$,
\end{definition}
also written
$f(\vec{a},\vec{b})=\langle\vec{a},\vec{b}\rangle=\vec{a}\cdot\vec{b}$.
The quantity $\sqrt{\langle \vec{a}, \vec{a} \rangle}$ defines the
Euclidean norm $\norm{\vec{a}}$ or magnitude of
$\vec{a}$.  Euclidean distance is the norm of the vector resulting
from the subtraction of two vectors,
$\norm{\vec{a}-\vec{b}}$. Euclidean distance is metric.

The dot product projects $\vec{a}$ onto $\vec{b}$ and scales the
result by the magnitude of $\vec{b}$. The scaling can lead to
unintuitive consequences. For example, two large identical vectors
have a larger dot product compared to two small identical vectors,
thus they would be considered ``more similar'' under this definition.

If magnitude is not important, $\vec{a}$ and $\vec{b}$ can be
normalized by $\vec{\hat{a}}=\vec{a}/\norm{\vec{a}}$ and
$\vec{\hat{b}}=\vec{b}/\norm{\vec{b}}$ so that they possess unit
magnitudes.  Then,

\begin{definition}[Cosine Similarity]
\label{def:cosine}
$f(\vec{a},\vec{b})=\langle\vec{\hat{a}},\vec{\hat{b}}\rangle$
\end{definition}
or $f(\vec{a},\vec{b})=
\frac{\langle\vec{a},\vec{b}\rangle}{\norm{\vec{a}}\norm{\vec{b}}}$.
Cosine similarity yields the angle between $\vec{a}$ and $\vec{b}$.


Arbitrary $p$-norms, $\norm{\vec{x}}_p=(|x_1|^p\dots |x_D|^p)^{1/p}$,
induce the Minkowski distance, generalizing Euclidean distance.

\begin{definition}[Minkowski]
The $p$-order Minkowski distance is
$d(\vec{a},\vec{b})=(\sum_{i=1}^n|a_i-b_i|^p)^{1/p}$,
\end{definition}
or $d(\vec{a},\vec{b})=\norm{\vec{a}-\vec{b}}_p$. All positive and
integer values of $p$ yield metrics.

Another generalization of Euclidean distance can be obtained by
applying a linear transformation over the vector space in order to
adjust the relative proximities of the feature vectors. The distance
of two vectors in the transformed space can be calculated using the
Mahalanobis formula.

\begin{definition}[Mahalanobis]
For any positive semi-definite matrix $\tens{M}$,
$d(\vec{a},\vec{b})=\sqrt{(\vec{a}-\vec{b})^\top\tens{M}(\vec{a}-\vec{b})}$.
\end{definition}

\subsubsection{Aggregate Scores}

Sometimes, a single real-world entity is represented by multiple
vectors in the vector collection. For example for facial recognition,
a face may be represented by multiple images taken from different
angles, leading to $m$ feature vectors $\vec{x}_1\dots\vec{x}_m$.

Given a query vector $\vec{q}$, finding the set of vectors which are
collectively most similar to $\vec{q}$ is called \emph{multi-vector
search}\footnote{Some VDBMSs such as \cite{weaviate} offer a
``hybrid'' search that is a multi-vector search over dense feature
vectors combined with sparse term vectors. The sparse vector is scored
separately, \textit{e.g.} by weighted term frequency, 
and then this score is combined with the feature vector similarity
score to yield a final aggregate score.\label{foot:hybrid}}. One way
of approaching this problem is to use an aggregate score that defines
how to combine individual scores $f(\vec{x}_1,\vec{q})\dots
f(\vec{x}_m,\vec{q})$ to yield a single value that can be
compared. Examples are the mean aggregate $1/m \sum_{i=1}^m
f(\vec{x}_i,\vec{q})$ and the weighted sum \cite{wang2021}.

\subsubsection{Learned Scores}

Some recent works \cite{abdelkader2019,zhang2020,meng2022} aim to find
a transformation, $\tens{M}$, in order to improve the quality of
similarity search over Mahalanobis distance.  Finding $\tens{M}$ is
one of the goals of \emph{metric learning}, and we point interested
readers to \cite{wang2015} for a survey of techniques.

%
%

\subsubsection{Discussion}

\hi{Score Selection.}
Many scores have been proposed over the years,
%
but how to select an appropriate score for a particular application
remains unclear.
Ideally, a score is selected so that query answers accurately reflect
the true semantic similarities between real-world entities represented
by the vectors. But so far, there are no principles guiding score
selection, and as a result, score selection tends to be based more on
informal rules distilled from experience rather than on rigorous
theory.

Many VDBMSs offer this as a choice to the user, and how to support
automatic score selection remains an open problem. We are aware of one
recent work \cite{wang2023} that dynamically adjusts the score based
on the query for social media content recommendation.

At a higher level, vector search is affected not only by the
similarity score but also by the nature of the embeddings and the
semantics of the query. A preliminary discussion on query semantics is
given in \cite{tagliabue2023}. Hence, we imagine that future
solutions will be more holistic, considering this problem from all
aspects beyond score selection.
For example, EuclidesDB \cite{euclid} allows users to conduct the same
search but over multiple embedding models and scores in order to
identify the most semantically meaningful settings.

\hi{Curse of Dimensionality.}
%
When $D$ grows beyond around 10 dimensions, and when the dimensions
are independent and identically distributed, the Euclidean distances
between the two farthest and two nearest vectors approach equality as
the variance nears zero \cite{beyer1999}.
The effect of this ``curse of dimensionality'' is that vectors become
indiscernible\footnote{A diagram of this phenomenon is given in
\cite{navarro2002}.}.

Attempts at avoiding the curse have led to using other Minkowski
distances, such as the Manhattan distance ($p=1$) or the Chebyshev
distance ($p=\infty$), in an effort to recover discernability
\cite{aggarwal2001}.
Fractional orders of $p<1$ have also been explored
\cite{mirkes2020} but so far, results are inconclusive.


\subsection{Queries and Operators}

Let $S\subset\mathbb{R}^D$ be a vector collection with $N$ members.
Let $\vec{q}$ be a $D$-dimensional query vector which may or may not
belong to $S$.

\emph{Data manipulation} queries aim to modify $S$. Given $S$,
$\vec{q}$, and a way to measure similarity (whether by $f$ or $d$),
\emph{vector search} queries aim to return a subset of $S$ where each
member of the subset satisfies criteria based on similarity to
$\vec{q}$. Different criteria lead to different types of queries.

%

To answer these queries, a VDBMS may use several basic operators in
addition to more sophisticated index operators that will be discussed
in Section~\ref{sec:indexing}.

\subsubsection{Data Manipulation Queries}

Data manipulation queries provide insert, update, and delete
mechanisms over $S$.
In a traditional data management system, data is manipulated
directly. But in a VDBMSs, feature vectors are proxies for the actual
entities, and they can be manipulated either directly or indirectly.
An \emph{embedding model} maps real-world entities (\textit{e.g.}
images) to feature vectors.

Under direct manipulation, users freely manipulate the values of the
vectors, and maintaining the model is the responsibility of the user.
This is the case for systems such as PASE \cite{yang2020} and
\texttt{pgvector} \cite{pgvec}.

For indirect manipulation, vectors are hidden from users. The vector
collection appears as a collection of entities, not vectors, and users
manipulate the entities.
The VDBMS is responsible for the model, which can be user-provided,
for example through a user-defined function (UDF) as in Vald
\cite{vald}, or selected from a menu of pre-trained models.
For example, Pinecone \cite{pine} supports a large number of
pre-trained \texttt{*2vec} models\footnote{\textit{E.g.}
\url{http://github.com/MaxwellRebo/awesome-2vec}} by connecting to
special providers\footnote{\textit{E.g.}
\url{http://huggingface.co}.} via REST API.

There is extensive literature on designing embedding models, and we
refer interested readers to \cite{pouyanfar2018}.

\subsubsection{Basic Search Queries}

\begin{figure}[t]
\centering
\includegraphics[width=0.45\textwidth]{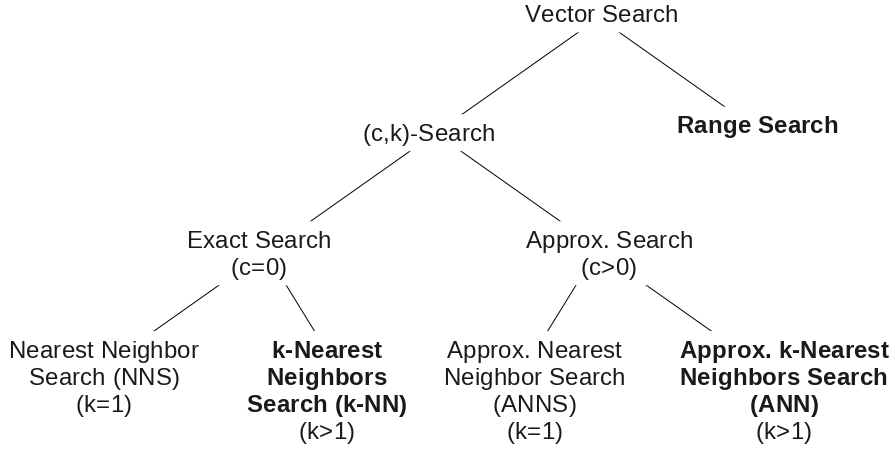}
\caption{Basic search queries.}
\label{fig:queries}
\end{figure}

Several types of search queries exist, but not all VDBMSs support all
types. The queries are shown in Figure~\ref{fig:queries}.


Search queries can be viewed as either similarity maximization or
distance minimization with respect to $\vec{q}$. We take the latter
for the following definitions.

\hi{$(c,k)$-Search Queries.}
Most VDBMSs support ``nearest neighbor'' queries, where the aim is to
retrieve vectors from $S$ that are physical neighbors of $\vec{q}$ in
the vector space. These queries may aim to return exact or approximate
nearest neighbors, and may also specify the number of neighbors to
return. We refer to these as $(c,k)$-search queries, where $c$
indicates the approximation degree and $k$ is the number of neighbors.

Out of these, most VDBMSs support the \emph{\underline{a}pproximate
$k$-\underline{n}earest \underline{n}eighbors} ({\ANN}) query, which
returns $k$ vectors from $S$ that are within a radius, centered over
$\vec{q}$, of $c$ times the
distance between $\vec{q}$ and its closest neighbor.

\begin{definition}[{\ANN}]
Find a $k$-size subset $S^\prime\subseteq S$ such that
$d(\vec{x}^\prime,\vec{q})\leq c(\min_{\vec{x}\in S}d(\vec{x},\vec{q}))$
for all $\vec{x}^\prime\in S^\prime$.
\end{definition}
If $k=1$, we call this query \emph{\underline{a}pproximate
\underline{n}earest \underline{n}eighbor \underline{s}earch}
(ANNS)\footnote{The terminology has become muddled over time. Early
efforts focused on exact search, with $c=1,k=1$, which was referred to as
``\underline{n}earest \underline{n}eighbor \underline{s}earch''
(NNS) \cite{indyk1998}. More recent efforts have focused on $c>1,k>1$,
referring to this query as ANN, ANNS, or other names.\label{foot:queries}}.

When $c=1$, we call this an exact query. The case $c=1,k=1$
corresponds to nearest neighbor search (NNS) (see
Note~\ref{foot:queries}), and when $c=1,k>1$, the query is called a
\emph{\underline{$k$}-\underline{n}earest \underline{n}eighbors}
({\KNN}) query.

%




We note that there is a large literature on the \emph{maximum inner
product search} (MIPS) problem, which is the NNS query but over inner
products.
We refer interested readers to \cite{teflioudi2016} for an
overview.

\hi{Range Queries.}
A range query is parameterized by a radius, $r$, instead of
the number of neighbors to return.

\begin{definition}[Range]
Find $\{ \vec{x} \in S \mid d(\vec{x},\vec{q})\leq r \}$.
\end{definition}
%

\subsubsection{Query Variants}

Some VDBMSs support variations on these basic query types, listed in
Table~\ref{tab:variants}.

\begin{table}[t]
\centering
\caption{Query variants.}
\label{tab:variants}
\begin{tabular}{ll}
\toprule
Variant
  & Sub-Variant \\
\midrule
Predicated
  & \\
Batched
  & \\
Multi-Vector
  & Multi-Query, Single Feature (MQSF) \\
  & Multi-Query, Multi-Feature (MQMF) \\
  & Single-Query, Multi-Feature (SQMF) \\
\bottomrule
\end{tabular}
\end{table}

\hi{Predicated Search Queries.}
In a predicated search query, or ``hybrid'' query, each vector is
associated with a set of attribute values, and a boolean predicate
over these values must evaluate to true for each record in the result
set\footnote{This type of query is called differently in different
systems. In some VDBMSs such as \cite{wei2020}, this is called a
``hybrid'' query, not to be confused with multi-vector search queries
mentioned in Note \ref{foot:hybrid}. In others such as \cite{qdrant},
this is called a ``filtered'' query.}. Here is an example:

\begin{example}
A hybrid {\KNN} query written in SQL is:
\begin{tcolorbox}
\begin{verbatim}
    select * from S where attr < c
    order by d(q) limit k;
\end{verbatim}
\end{tcolorbox}
\end{example}
In this example, \texttt{d} is a distance function parameterized by
the query \texttt{q}, and every member of the result set must satisfy
the conditions of being among the \texttt{k} nearest and of obeying
the predicate, \texttt{attr < c}.

\hi{Batched Queries.}
For batched queries, a number of queries are revealed to the system at
the same time, and the VDBMS can answer them in any order.

These queries are especially suited to hardware accelerated query
processing \cite{wang2021,johnson2021}.
%


\hi{Multi-Vector Queries.}
Some VDBMSs also support multi-vector search queries via aggregate
scores.

There are three possible sub-types: in multi-query single-feature
(MQSF) queries, the query is represented by multiple vectors, and
real-world entities are represented by single feature vectors; in
multi-query multi-feature (MQMF) queries, both the query and the
entities are represented by multiple vectors; and in single-query
multi-feature (SQMF) queries, only the entities are represented by
multiple vectors.
But so far, there is support for MQSF and SQMF queries
\cite{marqo,weaviate,milvus,wang2021,nuclia} but no support for MQMF
queries.

\subsubsection{Basic Operators}


%
It is possible to answer all $(c,k)$-search and range queries using
only projection.

\begin{definition}[Projection]
The projection of $S$ onto $g$ under query $\vec{q}$ yields $\pi_g(S)
= \{ g(\vec{x},\vec{q}) \mid \vec{x} \in S\}$.
\end{definition}

Projection onto $g=f$ or $g=d$ alone is sufficient for vector search
queries by projecting the full collection in time $O(\tau N)$, where
$\tau$ is the cost of $g$.

Typically $\tau$ is in $O(D)$, making projection by itself impractical
for VDBMSs where $N$ and $D$ are very large.
Thus, most VDBMSs rely on specialized index-based operators in
addition to projection.

\subsubsection{Discussion}

\hi{Query Accuracy and Performance.}
The search capability of a VDBMS is assessed by evaluating query
accuracy and performance.

To evaluate accuracy, precision and recall are often used. Precision
is defined as the ratio between the number of relevant results in the
result set over the size of the result set, and recall is defined as
the ratio between the number of retrieved relevant results over all
possible relevant results.
For example in a {\KNN} query, the precision is $k^\prime/|S^\prime|$,
where $k^\prime$ is the number of true nearest neighbors in
$S^\prime$, and the recall is $k^\prime/k$.

To evaluate performance, latency and throughput are used. Latency is
the amount of time it takes for a VDBMS to answer a query once it is
received, while throughput is the number of queries that are answered
per unit time, often reported as queries per second.

\hi{Theoretical Results.}
Similarity search has a long history, making it possible to summarize
some key theoretical results.
For low-dimensional NNS, many theoretical results are known\footnote{
Note that bounds on NNS also apply to {\KNN}.}.
Algorithms with $O(\log N)$ query times and $O(N)$ storage are known
for $D=1$ (\textit{e.g.} binary search trees) and $D=2$
\cite{lipton1980}. For the latter case, {\KDTREE}s \cite{bentley1975}
are particularly well known, and the query complexity is $O(\sqrt{N})$.
For $D\geq 3$, sub-linear search performance is much harder to
obtain. In the general case, {\KDTREE}s offer $O(DN^{1-1/D})$ query
complexity \cite{lee1977}, which tends toward $O(DN)$ as $D$ grows. On
the other hand, \cite{meiser1993} offers $O(D^{O(1)}\log N)$ query
complexity but requires super-polynomial $O(N^{O(D)})$ storage.
The modern belief is that even a fractional power of $N$ query
complexity cannot be obtained \emph{unless} storage cost is worse than
$N^{O(1)}D^{O(1)}$ \cite{andoni2018,rubinstein2018}.

%
%
Locality sensitive hashing (LSH) \cite{indyk1998} is perhaps the most
well understood {\ANN} algorithm.
The query complexity\footnote{Much effort has been on finding
``families'' of hash functions that can minimize $\rho$. For Hamming
distance, \cite{indyk1998} achieves $\rho\leq 1/c$, and for Euclidean
distance, \cite{andoni2006near,andoni2015} achieves $\rho\leq
1/c^2+o(1)$. These are both optimal \cite{odonnell2014}, although for
Euclidean distance the random projections family \cite{datar2004} is
more popular due to its simplicity. This family achieves $\rho< 1/c$.
%
%
These particular families are data-independent, but
by exploiting structural relationships within a vector collection,
\cite{andoni2015optimal} finds a family for Euclidean
distance which yields $\rho=1/(2c^2-1)$.}
is dominated by $N^\rho$ while the storage complexity is dominated by
$N^{1+\rho}$, where $0\leq\rho\leq 1$.
Recently, heuristic methods which lack guarantees have become popular.
%
There are encouraging analyses of graph-based techniques
\cite{laarhoven2018,prokhorenkova2020} but these lack
the maturity of LSH.

\subsection{Query Interfaces}


For query interfaces, native and NoSQL VDBMSs tend to rely on small APIs.
For example, Chroma \cite{chroma} offers a Python API with just nine
commands, including \texttt{add}, \texttt{update}, \texttt{delete}, and
\texttt{query}.

On the other hand, extended VDBMSs built over relational systems tend
to take advantage of SQL extensions. In \texttt{pgvector}
\cite{pgvec}, a {\KNN} or ANN query is expressed as:
\begin{tcolorbox}
\begin{verbatim}
    select * from items order by
    embedding <-> [3,1,2] limit 5;
\end{verbatim}
\end{tcolorbox}
The syntax \texttt{R <-> s} returns the Euclidean distance between all
the tuples of \texttt{R} and vector \texttt{s}, and other distance
functions are supported via other symbols.
If an ANN index is created over the \texttt{items} table, then this
query will return approximate results if the index is used for
execution.

Similarly, range queries are expressed using \texttt{where}:
\begin{tcolorbox}
\begin{verbatim}
    select * from items where
    embedding <-> [3,1,2] < 5;
\end{verbatim}
\end{tcolorbox}


\section{Indexing}
\label{sec:indexing}

While all $(c,k)$-search and range queries can be answered by
comparing $\vec{q}$ with each of the $N$ vectors, the complexity of
this brute-force approach is $O(DN)$, prohibitive for large $D$ and
$N$.

Instead, vector indexes speed up queries by minimizing the number of
comparisons. This is achieved by partitioning $S$ so that only a small
subset is compared, and then arranging the partitions into data
structures that can be easily traversed.

Unlike typical attributes, vectors are not obviously sortable nor can
they be easily categorized. To achieve high accuracy, these indexes
rely on novel techniques which we refer to as
\emph{randomization}, \emph{learned partitioning}, and \emph{navigable
partitioning}.
The large physical size of vectors also leads to use of compression,
namely a technique called \emph{quantization}, as well as \emph{disk
resident} designs. Additionally, the need to support predicated
queries has led to special \emph{hybrid operators} for indexes,
which we will discuss in Section~\ref{sec:opt}.

\hi{Partitioning Techniques.}
\begin{itemize}
\item %
\emph{Randomization.}
Randomization aims to exploit probability amplification of multiple
independent events,
%
allowing indexes to better discriminate truly similar vectors from
dissimilar ones.
\item %
\emph{Learned Partitioning.}
Learning-based techniques aim to uncover an internal structure of $S$
so that it can be partitioned along this structure. These techniques
can be supervised or unsupervised.
\item %
\emph{Navigable Partitioning.}
Instead of fixating on absolute partitions, navigable indexes are
designed so that different regions of $S$ can be easily traversed.
\end{itemize}
Some partitioning strategies are \emph{data independent}, where the
rules are the same for any data distribution. But the majority are
\emph{data dependent}. If updates to $S$ alter its distribution, then
indexes based on data dependent strategies may eventually become
unbalanced over time. In many cases, this can only be resolved by
rebuilding the index.

\hi{Storage Techniques.}
\begin{itemize}
\item %
\emph{Quantization.} A quantizer maps a vector onto a more
space-efficient representation. Quantization is usually lossy, and the
aim is to minimize information loss while simultaneously minimizing
storage cost.
\item %
\emph{Disk-Resident Designs.} Compared to memory resident indexes
which only minimize the number of comparisons, disk resident indexes
additionally aim to minimize the number of retrievals.
\end{itemize}

Below, we examine the main techniques for some common indexes. One
particular index may use a combination of techniques, and so we
classify indexes based on their structure and then point out
which techniques are used in which index.
There are three basic structures: \emph{tables} divide $S$ into
buckets containing similar vectors; \emph{trees} are a nesting of
tables; \emph{graphs} connect similar vectors with virtual edges that
can then be traversed.
%
%
%
All of these structures are capable of achieving high query accuracy
but with different construction, search, and maintenance
characteristics.

\subsection{Tables}

\begin{table}[t]
\centering
\caption{Representative table-based indexes.}
\label{tab:tables}
\begin{tabular}{lll}
\toprule
Type
  & Index
  & Hash Function \\
\midrule
LSH
  & E$^2$LSH \cite{datar2004}
  & Rand. hyperplanes \\

  & \texttt{IndexLSH}
  & Rand. bits \\

  & FALCONN \cite{andoni2015}
  & Rand. balls \\
L2H
  & SPANN \cite{chen2021}
  & Nearest centroid \\
Quant.
  & \texttt{SQ}
  & Nearest discrete value \\

  & \texttt{PQ}
  & Nearest centroid product \\

  & IVFSQ
  & Nearest centroid \\

  & IVFADC \cite{jegou2011product}
  & Nearest centroid \\
\bottomrule
\end{tabular}

\end{table}

The main consideration for table-based indexes is the design of the
bucketing hash function.
The most popular table-based indexes for VDBMSs tend to use
randomization and learned partitions, as shown in
Table~\ref{tab:tables}.
For randomization, techniques based on LSH \cite{andoni2008,%
andoni2015,andoni2018,datar2004,indyk1998,leskovec2014,lv2007}
are popular due to robust error bounds.
For learned partitions, learning-to-hash (L2H) \cite{wang2018}
directly learns the hash function, and  indexes based on quantization
\cite{gray1984,jegou2011product,matsui2018,sivic2003,%
wang2020,wei2020} typically uses $k$-means \cite{berg2008} to learn
geometric clusters of similar vectors.

All of these indexes have similar construction and search
characteristics.
For construction, each vector is hashed into one of the buckets. The
total complexity is typically $O(DN^{1+\epsilon})$, $0\leq\epsilon\leq
1$. For LSH, each vector is hashed multiple times, leading to
$\epsilon>0$. For quantization-based approaches, $k$-means multiplies
the complexity by a constant factor. 
For search, $\vec{q}$ is hashed onto a key and then the corresponding
bucket is scanned. Hashing is generally on the order of $O(D)$.
Usually only a small fraction of $S$ is scanned, yielding a complexity
of about $O(DN^\epsilon)$, including the cost of hashing and bucket
scan.

\subsubsection{Locality Sensitive Hashing}

Locality sensitive hashing \cite{indyk1998,leskovec2014} provides
tunable performance with error guarantees, but it can require
high redundancy in order to boost accuracy, increasing query and
storage costs relative to other techniques.

In a ``family'' of hash functions
$H=\{h:S\rightarrow U\}$,
if $d(\vec{x},\vec{q})\leq r_1$, then
$\Pr_H(h(\vec{x})=h(\vec{q}))\geq p_1$, and if $d(\vec{x},\vec{q})\geq
r_2$, then $\Pr_H(h(\vec{x})=h(\vec{q}))\leq p_2$, for any $r_1$, $r_2$,
 $\vec{x}\in S$, and $\vec{q}$.
A tunable family $G=\{g:S\rightarrow U^K\}$ is derived by letting
$g(\vec{x})$ return the concatenation of $h_i(\vec{x})$ for $i$
between $1$ and some constant $K$.

%
%

The table is constructed by hashing each $\vec{x}\in S$ into each of
the $L$ hash tables using $g_1\dots g_L$.
Typically, $L$ is set to $L=O(1/p_1^K)$ with $K$ set to $\lceil
\log_{1/p_2} N \rceil$ \cite{andoni2018}. The exact value depends on
the accuracy and performance needs of the application, and some sample
curves are shown in \cite{andoni2008}.
Letting $\rho=\log(1/p_1)/\log(1/p_2)$ yields $L=O(N^\rho/p_1)$. The
storage complexity is $O(LDN)$ which is $O(DN^{1+\rho})$ after
substitution. In the practical case where $p_1>p_2$, the value of
$\rho$ is between $0$ and $1$.

\begin{figure}[t]
\centering
\includegraphics[width=0.30\textwidth]{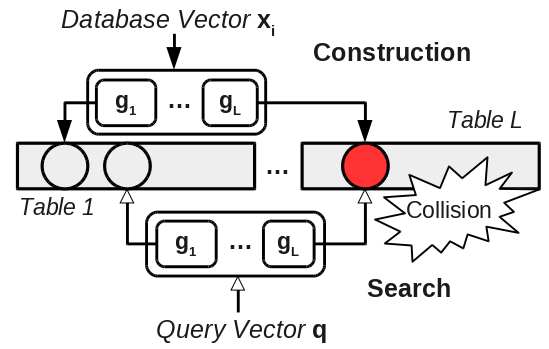}
\caption{Constructing and searching an LSH index.}
\label{fig:lsh}
\end{figure}

When a query appears, it is hashed using the $L$ hash functions
sampled from $G$, and collisions are kept as candidate neighbors. The
candidates are then re-ranked or discarded based on true distances to
$\vec{q}$. Figure~\ref{fig:lsh} illustrates this procedure.
The query complexity is dominated by the $L$ hash evaluations, which
is $O(DN^\rho)$.

%

When $r_1$ is set to $\min_{\vec{x}\in S}d(\vec{x},\vec{q})$ and $r_2$
is set to $cr_1$, the $c$ guarantee is relative to the minimum
distance. This is useful when the query is static across the workload,
but is is hard to generalize over dynamic online queries.  Hence for
an index designed around some given hash family, not all queries may
have similar candidate sets, making it hard to control precision and
recall. Multi-probe LSH \cite{lv2007} is one attempt at addressing
this issue by scanning multiple buckets at a time, thereby spreading
out the search.

We mention a few popular LSH schemes. The first two are \emph{data
independent} and require no rebalancing.
\begin{itemize}
\item%
\emph{E$^2$LSH.} Each $g$ is an $O(D)$ projection onto a random
hyperplane. This achieves $\rho<1/c$
\cite{datar2004}.
\item %
\emph{\texttt{IndexLSH}.} This scheme is based on binary projections
and is provided by Faiss \cite{faiss}.
\end{itemize}
There have also been efforts at designing \emph{data dependent} hash
families to yield lower $\rho$.
\begin{itemize}
\item %
\emph{FALCONN.} Implements an LSH hash family based on spherical LSH
\cite{andoni2015}. The dataset is first projected onto a unit ball and
then recursively partitioned into small overlapping spheres. The
$\rho$ value is $1/(2c^2-1)$.
\end{itemize}
Other families are given in \cite{andoni2018}.

\subsubsection{Learning to Hash}

Learning-based techniques aim to directly learn suitable mappings
without resorting to hash families.
For example, spectral hashing \cite{weiss2008} designs hash functions
based on the principal components of the similarity matrix of $S$. In
\cite{salakhutdinov2007}, hash functions are modeled using neural
networks. In SPANN \cite{chen2021}, vectors are hashed to the nearest
centroid following $k$-means.
Several techniques for a disk resident table-based index are also
given.

These techniques tend to require lengthy training and are sensitive to
out-of-distribution updates, and they are not widely supported in
VDBMSs. We point readers to \cite{wang2018} for a survey of
techniques.

\subsubsection{Quantization}

One of the main criticisms of LSH is that the storage cost can be
large due to the use of multiple hash tables. For in-memory VDBMSs,
large memory requirements can be impractical. While some efforts
aim at reducing these storage costs \cite{zhang2016}, other
efforts have targeted vector compression using quantization
\cite{gray1984,sivic2003,jegou2011product}.
%

Many of these techniques use $k$-means centroids as hash
keys\footnote{For a discussion, see
\cite{lloyd1982,gray1984,gersho1991}.}. A large $K$ number of
centroids can modulate the chance of collisions, keeping buckets
reasonably small and speeding up search. Normally, $k$-means is
terminated once a locally optimal set of centroids is found, with
complexity $O(DN\cdot K)$ per iteration.

\begin{figure}[t]
\centering
\includegraphics[width=0.4\textwidth]{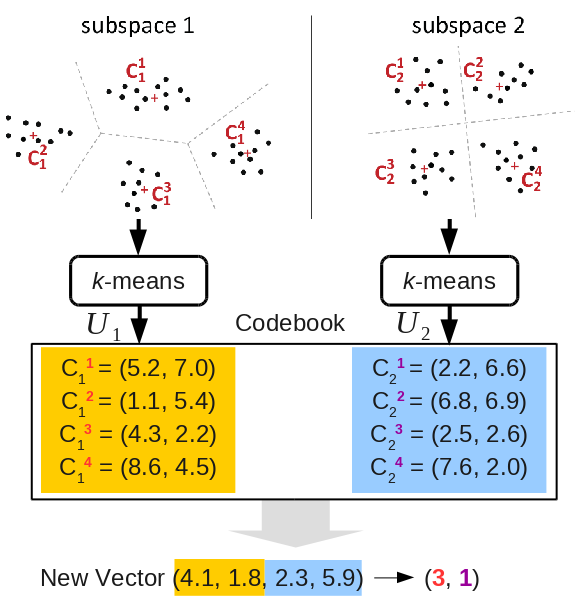}
\caption{In this example from \cite{wang2020}, each 4-dimensional
vector is divided into two 2-dimensional sub-spaces.}
\label{fig:wang}
\end{figure}

But large $K$ make $k$-means expensive.
\emph{Product quantization} exploits the fact that the cross product of $m$
number of $(D/m)$-dimensional spaces is a space of $D$ dimensions, so
that by setting $U=\prod_{j=1}^m U_i$, then $U\in\mathbb{R}^D$ when
$U_j\in\mathbb{R}^{D/m}$. This means that to yield a count of $K$
centroids, only $K^{1/m}$ centroids need to be found per
$U_j$. Moreover as each $U_j$ belongs to a lower dimensional space,
the running time of $k$-means per $U_j$ is reduced.
The new complexity is $O(m)O(\frac{D}{m}NK^{1/m}i)$.

Each $U_j$ is constructed via $k$-means over the collection of
sub-vectors $\{(x_i)^{jD/m}_{i=(j-1)D/m+1}\mid \vec{x}\in
S\}$\footnote{The notation $(x_i)^D_{i=1}$ expands to $x_1x_2\dots
x_D$.}, and the set of all $U_j$ is known as the ``codebook''.
Vector $\vec{x}$ is then quantized by splitting it into $m$
sub-vectors, $\vec{x}^\prime_j$, finding the nearest centroid in $U_j$
to $\vec{x}^\prime_j$ for each $j\in 1\dots m$, and then concatenating
these centroids.
Each vector is thus stored using $m\log_2(D/m)$ bits, and
the time complexity is $O(m)O(DK^{1/m})$. An example is shown
in Figure~\ref{fig:wang}.

Various techniques such as Cartesian $k$-means \cite{norouzi2013},
optimized PQ (OPQ) \cite{ge2013}, hierarchical quantizers
\cite{yandex2016}, and anisotropic quantizers \cite{guo2020} have been
developed based on this idea, offering around 60\% better recall in
the best cases at the cost of additional processing. A survey of
these techniques is given in \cite{matsui2018}. The storage cost can
also be reduced by constant factors \cite{wang2020}.

\begin{figure}[t]
\centering
\includegraphics[width=0.48\textwidth]{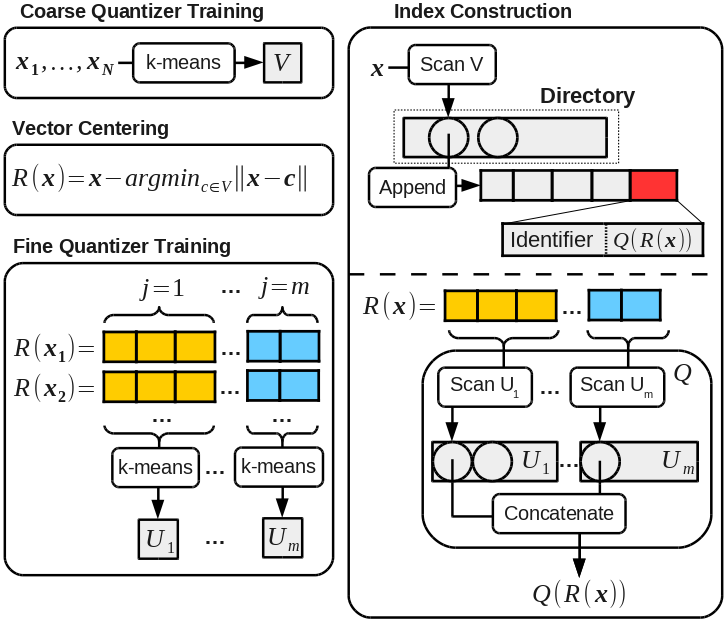}
\caption{Construction of an IVFADC index.}
\label{fig:pq-construction}
\end{figure}

We list some quantization-based indexes.
\begin{itemize}
\item %
\emph{Flat Indexes.}
Faiss \cite{faiss} supports a number of ``flat'' indexes where each
vector is directly mapped onto its compressed version, without any
bucketing.
The standard quantizer index, \texttt{SQ}, performs a bit-level
compression, for instance by mapping 64-bit doubles onto 32-bit
floats\footnote{Also called ``lattice'' quantization, see
\cite{agrell2023}.}.
The \texttt{PQ} index directly maps each vector onto its PQ code.
\item %
\emph{IVFSQ.}
For IVFSQ, the vectors are compressed using \texttt{SQ} and
bucketed to their nearest centroid.
\item%
\emph{IVFADC.}
Training a PQ quantizer over $S$ can still be time consuming. To
reduce this cost, IVFADC first buckets vectors using $k$-means over a
small number of centroids, and then trains a PQ quantizer by sampling
a few vectors from each of the buckets. To allow a single quantizer to
apply to all the buckets, each vector $\vec{x}$ is normalized by
subtracting from its bucket key, resulting in a ``residual''
vector $R(\vec{x})$ which is then used to train the
quantizer. The full workflow is shown in Figure~\ref{fig:pq-construction}.
%
%
During search, query $\vec{q}$ is directly compared against the
quantized vectors in the bucket that $\vec{q}$ maps onto. As
$\vec{q}$ itself is not quantized, the comparison is referred to as an
``asymmetric distance computation'' (ADC).
\end{itemize}

For IVFADC, many distance calculations are likely to be repeated
during bucket scan since many vectors may share the same PQ
centroids. These calculations can be avoided by first computing
$\norm{\overline{\vec{q}_j}-\vec{c}}^2$ for all $\vec{c}\in U_j$ and
for all $j\in 1\dots m$, where $\overline{\vec{q}_j}$ is the $j$th
sub-vector of $\vec{q}$ \cite{matsui2018}.
This preprocessing step takes $O(m)O(\frac{D}{m}K^\prime)$, where
$K^\prime$ is the number of centroids in $U_j$.
But afterwards, ADC can be performed using just $m$ look-ups, reducing
bucket scan from $O(DN)$ to $O(mN)$.

\begin{example}
\label{ex:adc}
Below is the ADC look-up table when $U$ is divided into $m$ subsets,
and where each subset contains $K^\prime$ centroids. Here,
$\overline{\vec{q}_j}$ is the $j$th sub-vector of query $\vec{q}$, and
$\vec{c}^j_i$ is the $i$th centroid in the $j$th subset of $U$.
\begin{equation*}
\overbrace{
\begin{matrix}
        d(\overline{\vec{q}_1},\vec{c}^1_1)&
\cdots &d(\overline{\vec{q}_1},\vec{c}^1_{K^\prime}) \\
\vdots & \ddots & \vdots \\
        d(\overline{\vec{q}_m},\vec{c}^1_1)&
\cdots &d(\overline{\vec{q}_m},\vec{c}^1_{K^\prime})
\end{matrix}}^{U_1}
,\cdots,
\overbrace{
\begin{matrix}
        d(\overline{\vec{q}_1},\vec{c}^m_1)&
\cdots &d(\overline{\vec{q}_1},\vec{c}^m_{K^\prime}) \\
\vdots & \ddots & \vdots \\
        d(\overline{\vec{q}_m},\vec{c}^m_1)&
\cdots &d(\overline{\vec{q}_m},\vec{c}^m_{K^\prime})
\end{matrix}}^{U_m}
\end{equation*}
\end{example}

We mention one other technique.
In AnalyticDB-V \cite{wei2020}, each bucket is further divided into
finer sub-buckets in order to avoid accessing multiple full
buckets. The resulting structure is called a ``Voronoi Graph Product
Quantization'' (VGPQ) index.

\subsection{Trees}


For tree-based indexes, the main consideration is the design of the
splitting strategy used to recursively split $S$ into a search tree.

A natural approach is to split based on distance. The main techniques
includes pivot-based trees \cite{chen2022,chen2022survey}, such as
VP-tree \cite{yianilos1993} and M-tree \cite{ciaccia1997}, $k$-means
trees \cite{muja2009}, and trees based on deep learning
\cite{li2023}. Other basic techniques are described in
\cite{sellis1997}.
But while these trees are effective for low $D$, they suffer from the
curse of dimensionality when applied to higher dimensions.

High-$D$ tree-based indexes tend to rely on randomization for
performing node splits.
In particular, ``Fast Library for ANN'' (FLANN) \cite{flann,muja2009}
combines randomization with learned partitioning via principal
component analysis (PCA), extending the PKD-tree technique from
\cite{silpa2008}, and ``ANN Oh Yeah'' (ANNOY) \cite{annoy} is similar
to the random projections tree (RPTree) from
\cite{dasgupta2008,dasgupta2013}. These trees are summarized in
Table~\ref{tab:trees}.

%

The generic tree construction algorithm, from \cite{dasgupta2008}, is
restated below:
\begin{algorithmic}[0]
\Procedure{MakeTree}{$S$}
\If {$|S|\leq \tau$}
\Return Leaf
\EndIf
\State $\text{Rule}\leftarrow\text{ChooseRule}(S)$
\State $\text{LeftTree}\leftarrow\text{MakeTree}(\{\vec{x}\in S\mid \text{Rule}(\vec{x})\text{ is true}\})$
\State $\text{RightTree}\leftarrow\text{MakeTree}(\{\vec{x}\in S\mid \text{Rule}(\vec{x})\text{ is false}\})$
\State\Return $(\text{Rule}, \text{LeftTree}, \text{RightTree})$
\EndProcedure
\end{algorithmic}
The complexity is characteristically $O(DN\log N)$, ignoring any
preprocessing costs. More precise bounds for several trees are given
in \cite{ram2019}.

Most trees are able to return exact query results by performing
backtracking, where neighboring leaf nodes are also checked during the
search.
However, this is inefficient \cite{weber1998}, and they are more often
used for returning approximate results using \emph{defeatist search}
\cite{dasgupta2013}. In this procedure, the tree is traversed down to
the leaf level, and all vectors within the leaf covering $\vec{q}$ are
returned immediately as the nearest neighbors. There is no
backtracking, and the complexity is $O(D\log N)$.

For maintenance, insertions require $O(D\log N)$ on average and
$O(DN)$ in the worst case. But as node splits are determined during
construction, there is no obvious way to rebalance the nodes after a
number of out-of-distribution insertions. Self-balancing trees exist
for scalar data \cite{adelson1962} but not for high-dimensional
vectors.

\begin{table}[t]
\centering
\caption{Representative tree-based indexes.}
\label{tab:trees}
\begin{tabular}{lll}
\toprule
Index
  & Splitting Plane
  & Splitting Point \\
\midrule
{\KDTREE} \cite{bentley1975}
  & Axis parallel
  & Median \\
PKD-tree \cite{silpa2008}
  & Principal dim.
  & Median \\
FLANN \cite{muja2009}
  & Random principal dim.
  & Median \\
RPTree \cite{dasgupta2008,dasgupta2013}
  & Random plane
  & Median + offset \\
ANNOY \cite{annoy}
  & Random plane
  & Random median \\
\bottomrule
\end{tabular}

\end{table}

\subsubsection{Non-Random Trees}

Many high-$D$ trees derive from {\KDTREE}, which splits along
medians while rotating through dimensions:

\begin{algorithmic}[0]
  \Procedure{ChooseRule}{$S$}
  \State $i\leftarrow l\mod D$ where $l$ is the current depth
  \State $\text{Rule}\coloneqq(x_i\leq \text{median}(\{y_i\mid \vec{y}\in S\}))$
  \State\Return Rule
  \EndProcedure
\end{algorithmic}
This rule has the effect of fixing the splitting planes parallel to
the dimensional axes \cite{dasgupta2008,ram2019}.

\subsubsection{Random Trees}

If certain dimensions explain the variance more than others,
then the \emph{intrinsic} dimensionality\footnote{A formal definition
is given in \cite{dasgupta2008}.} is lower than $D$. But in this case,
{\KDTREE} is unable to partition along these dimensions, leaving it
susceptible to the curse of dimensionality. This limitation has led to
the discovery of more adaptive splitting strategies.


%

\hi{Principal Component Trees.}
A principal component tree is a {\KDTREE} that is constructed by first
rotating $S$ so that the axes are aligned with the principal
components of $S$. The principal dimensions need to be found
beforehand using principal component analysis (PCA). The complexity of
this step is $O(D^2N+D^3)$.
\begin{itemize}
\item %
\emph{PKD-tree.}
In PKD-tree, the splitting plane is selected by rotating through the
principal dimensions, similar to in vanilla {\KDTREE}
\cite{silpa2008}.
\item %
\emph{FLANN.}
Instead of rotating through the dimensions, FLANN \cite{muja2009}
splits along random principal dimensions.
\end{itemize}

\hi{Random Projection Trees.}
On the other hand, random splitting planes can be used to adapt to the
intrinsic dimensionality without expensive PCA.
\begin{itemize}
\item %
\emph{RPTree.}
RPTree \cite{dasgupta2008,dasgupta2013} extends the idea of
randomly rotated trees explored in \cite{vempala2012} by introducing
random splits in addition to random splitting planes.
The principle follows from spill trees \cite{liu2004}, where
partitions are allowed to overlap. In RPTree, perturbed median splits
simulate the effects of overlapping splits but with less storage
cost. The splitting rule is \cite{dasgupta2013}:
\begin{algorithmic}[0]
  \Procedure{ChooseRule}{$S$}
  \State $\vec{u}\leftarrow\text{A vector from the unit sphere}$
  \State $\beta\leftarrow\text{A number from }[0.25, 0.75]$
  \State $v\leftarrow\text{The }\beta
    \text{ fractile of }\pi_{\vec{u}}(S)$
  \State $\text{Rule}\coloneqq(\vec{x}\cdot\vec{u}\leq v)$
  \State\Return Rule
  \EndProcedure
\end{algorithmic}
The values of $\vec{u}$, $\beta$, and $\vec{v}$ are uniformly chosen
at random. The operation $\pi_{\vec{u}}(S)$ performs a projection of
$S$ onto $\vec{u}$. Variable $v$ represents the perturbed
median split, with $\beta=0.5$ yielding the true median.
\item %
\emph{ANNOY.}
Instead of splitting on the $\beta$ fractile, ANNOY splits along the
median of two random values sampled from $\pi_{\vec{u}}(S)$, which is
simpler to compute.
\end{itemize}

Several theoretical results\footnote{%
The rate at which RPTree fails to return a result set containing the
true nearest neighbor, $\vec{x}^\prime$, to query $\vec{q}$ is
bounded by the ``potential'' of the query over $S$, defined as
$1/(N-1)\sum_{i=2}^N
\norm{\vec{q}-\vec{x}^\prime}^2/\norm{\vec{q}-\vec{x}_i}^2$. A detailed
proof is available in \cite{dasgupta2013}.}\footnote{%
During search, query $\vec{q}$ must be projected onto each
unit vector at every level during traversal, leading to a search
complexity of $O(D\log N)$. In \cite{ram2019}, this is improved to
$O(D\log D+\log N)$ by using a circular rotation which can be applied
in $O(D\log D)$ time and achieves a similar effect as random
projections.
The trinary projection (TP) tree introduced in \cite{wang2014}
similarly targets expensive $O(D)$ projections. Instead of projecting
onto random or principal vectors, the splitting strategy partitions
onto principal trinary vectors, which are vectors consisting of only
$-1$, $0$, or $1$. The principal trinary vectors can be approximated
in $O(D)$ time. The search complexity remains $O(D\log N)$ but with
smaller constant factors.
} are known for RPTree, but it is not clear if these apply to ANNOY.

\begin{figure}[t]
\centering
\includegraphics[width=0.4\textwidth]{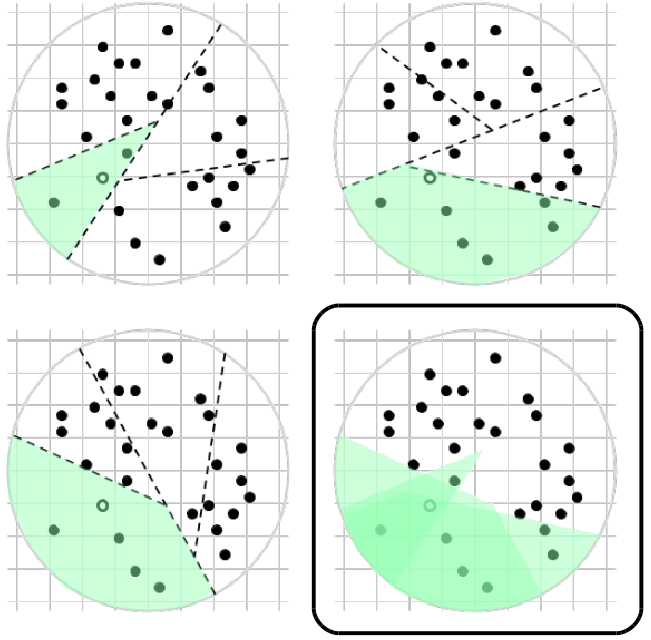}
\caption{As shown in \cite{ram2019}, a forest of RPTrees can be used to
improve recall by merging their individual result sets (bottom right)
to the query (empty circle).}
\label{fig:forest}
\end{figure}

As shown in Figure~\ref{fig:forest}, a forest of random trees can be
used to improve recall.
We mention that RPTree incurs a storage overhead of $O(DN)$ compared
to {\KDTREE} and FLANN due to storing the $D$-dimensional projection
vectors, and this cost can be substantial for in-memory forests. In
\cite{keivani2018}, several techniques are introduced to reduce this
cost to $O(D\log N)$, for example by combining projections across the
trees in a forest.


\subsection{Graphs}

\begin{figure}[t]
\centering
\includegraphics[width=0.45\textwidth]{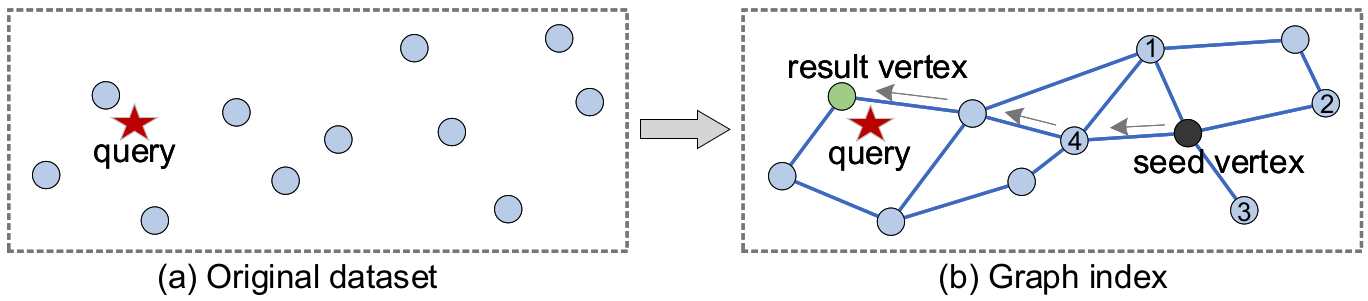}
\caption{In this example from \cite{wang2021comprehensive}, a graph
is overlayed on top of the original dataset (a). Search begins from
a ``seed'' vertex and is guided along the edges until it reaches
the nearest node (vector) to the query (b).}
\label{fig:wang2021}
\end{figure}

A graph-based index is constructed by overlaying a graph on top of the
vectors in $\mathbb{R}^D$ space, so that each node $v_i$ is positioned
over the vector $\vec{x}_i$ within the space. This induces distances
over the nodes, $d(v_i,v_j)=d(\vec{x}_i,\vec{x}_j)$, which are then
used to guide vector search along the edges. An example is
shown in Figure~\ref{fig:wang2021}.

The main consideration for these indexes is edge selection, in other
words deciding which edges should be included during graph
construction.

Graph-based indexes encapsulate all the partitioning techniques.
Many graphs rely on random initialization or random sampling during
construction. The $k$-nearest neighbor graph (KNNG)
\cite{eppstein1997,paredes2005} associates each vector with its $k$
nearest neighbors through an iterative refinement process similar to
$k$-means and which we consider to be a form of unsupervised learning.
Other graphs, including monotonic search networks (MSNs)
\cite{dearholt1988} and small-world (SW) graphs
\cite{malkov2014,malkov2020}, aim to be highly navigable, but differ
in their construction. The former tend to rely on search trials that
probe the quality of the graph \cite{harwood2016,fu2019,%
subramanya2019} while the latter use a heuristic procedure which we
refer to as ``one-shot refine''.
Table~\ref{tab:graphs} shows several graph indexes.

\begin{table}[t]
\centering
\caption{Representative graph-based indexes.}
\label{tab:graphs}
\begin{tabular}{llll}
\toprule
Type
  & Index
  & Initialization
  & Construction \\
\midrule
KNNG
  & KGraph \cite{dong2011}
  & Random KNNG
  & Iterative refine \\

  & EFANNA
  & Random trees
  & Iterative refine \\
MSN
  & FANNG \cite{harwood2016}
  & Empty graph
  & Random trial \\

  & NSG \cite{fu2019}
  & Approx. KNNG
  & Fixed trial \\

  & Vamana \cite{subramanya2019}
  & Random graph
  & Fixed trial \\
SW
  & NSW \cite{malkov2014}
  & Empty graph
  & One-shot refine \\

  & HNSW \cite{malkov2020}
  & Empty graph
  & One-shot refine \\
\bottomrule
\end{tabular}

\end{table}


Graph-based search techniques have been shown to perform well in
practice \cite{li2020}, and recent analytical results suggest that
asymptotic performance approaches the $N^\rho$ and $N^{1+\rho}$ limits
achieved by LSH, albeit with $\rho=c^2/(2c^2-1)$ but smaller constant
factors \cite{laarhoven2018,prokhorenkova2020}.
An experimental comparison of graph-based techniques is available in
\cite{wang2021comprehensive}.

\subsubsection{$k$-Nearest Neighbor Graphs}

In a KNNG, each node $v_i$ is connected to $k$ nodes representing the
nearest neighbors to $\vec{x}_i$ \cite{eppstein1997}.
For batched queries, $\vec{q}$ can be considered as a member of $S$,
and a KNNG built over $S$ allows exact {\KNN} search in $O(1)$ time
through a simple look-up.

A KNNG can also be used to answer interactive queries, where
$\vec{q}\notin S$. The basic idea is to recursively select node
neighbors that are nearest to $\vec{q}$, starting from initial nodes,
and add them into the top-$k$ result set. The search complexity
depends on the number of iterations before the result set converges.
The search can start from multiple initial nodes, and if there are no
more node neighbors to select, it can be restarted from new initial
nodes \cite{wang2012query}.

A KNNG can be exact or approximated with a technique which we refer to
as ``iterative refine''.

\hi{Exact.}
An exact KNNG can be constructed by performing a brute force search
$N$ number of times, giving a total complexity of
$O(DN^2)$. Unfortunately, there is little hope for improvement, as it
is believed that the complexity is bounded by $N^{2-o(1)}$
\cite{williams2018}. An $O(N\log N)$ algorithm is given in
\cite{vaidya1989} but with a constant factor that is $O(D^D)$. The
algorithm in \cite{paredes2006} achieves an empirical complexity of
$O(N^{2-\epsilon})$, where $0<\epsilon<1$. This finding suggests the
existence of efficient practical algorithms, despite the
worst-case bounds.

\hi{Iterative Refine.}
An approximate KNNG can be obtained by iteratively refining an initial
graph. We give two examples.
\begin{itemize}
\item %
\emph{NN-Descent.}
The NN-Descent (KGraph) method \cite{dong2011} begins with a random
KNNG and iteratively refines it by examining the neighbors of the
neighbors of each node $v_i$, replacing edges to $v_i$ with edges to
these second-order neighbors that are closer.
When the dataset is growth restricted\footnote{A growth restricted
dataset is one where the number of neighbors of each node is bounded
by a constant as the radius about the node expands.}, then each
iteration is expected to halve the radius around each node and its
farthest neighbor. This property leads to fast convergence, with
empirical times on the order of $O(N^{2-\epsilon})$ for
$0<\epsilon<1$.
\item %
\emph{EFANNA.}
Instead of starting from a random KNNG,
EFANNA\footnote{\url{http://arxiv.org/abs/1609.07228}} uses a forest
of randomized {\KDTREE}s to build the initial KNNG. Doing so is shown to
lead to higher recall and faster construction, as it can quickly
converge to better local optima. 
A similar approach is taken in \cite{wang2012} but where the
tree is constructed via random hyperplanes.
\end{itemize}

\subsubsection{Monotonic Search Networks}

A KNNG is not guaranteed to be connected. Disconnected components
complicates the search procedure for online queries by requiring
restarts to achieve high accuracy
\cite{paredes2005,wang2012query}. But by adding certain edges so that
the graph is connected, it becomes possible to follow a single path
beginning from any initial node and arriving at the nearest neighbor
to $\vec{q}$.

A search path $v_1\dots v_m$ is \emph{monotonic} if
$d(v_i,\vec{q})>d(v_{i+1},\vec{q})$ for all $i$ from $1$ to $m-1$.  An
MSN is a graph where the search path discovered by a ``best-first''
search, in which the neighbor of $v_i$ that is nearest to $\vec{q}$ is
greedily selected, is always monotonic. This property implies a
monotonic path for every pair of nodes in the graph and that the
graph is connected.

The search complexity depends on the sum of the out-degrees over the
nodes in the search path.
If there are few total edges in the graph, then the search complexity
is likely to be small. The minimum-edge MSN that guarantees exact NNS
is believed to be the Delaunay triangulation \cite{navarro2002}. But
constructing a triangulation requires at least $\Omega(N^{\lceil D/2
\rceil})$ time \cite{edelsbrunner1996}, impractical for large $N$ and
$D$. As a result, several approximate methods have been developed, but
these necessarily sacrifice the search guarantee.

In the early work by \cite{dearholt1988}, an MSN is constructed in
polynomial time by refining a sub-graph of the Delaunay triangulation
called the relative neighborhood graph (RNG), introduced in
\cite{toussaint1980} and later reviewed in \cite{jaromczyk1992}. The
RNG itself is not monotone, but it can be constructed in
$O(DN^{2-o(1)}\log^{1-o(1)} N)$ time under $\mathbb{R}^D$
Euclidean distance \cite{su1991}.
But the $N^{2-o(1)}$ term makes this approach impractical for large
$N$.

Instead of repeatedly scanning the nodes, \emph{search trials} can be
used to probe the quality of the graph. Depending on the path taken by
best-first search, new edges are added so that a monotonic path exists
between source and target. The algorithm is shown below.

\begin{algorithmic}[0]
\Procedure{MakeMSN}{$S$}
\State $G\leftarrow\text{InitializeGraph(S)}$
\Repeat
\State $(s,t)\leftarrow\text{ChooseSourceTargetPair}(S)$
\State $P\leftarrow\text{GetSearchPath}(G,(s,t))$
\State $\text{UpdateOutNeighbors}(G,t,P)$
\For {each $p\in P$}
\State $\text{UpdateOutNeighbors}(G,p,t)$
\EndFor
\Until{$\text{Terminate}()$}
\State \Return $G$
\EndProcedure
\end{algorithmic}
For InitializeGraph, some indexes begin with an empty graph
\cite{harwood2016}, random graph \cite{subramanya2019}, or
approximate KNNG \cite{fu2019}. Simple graphs can be initialized
quickly but more complex graphs may offer better quality. For
ChooseSourceTargetPair, one way is to select random pairs
\cite{harwood2016}, while another is to designate a node as the source
for all search trials \cite{subramanya2019,fu2019}. We refer to
these techniques as \emph{random} and \emph{fixed} trials, respectively.

\hi{Random Trial.}
Indexes based on random trials are constructed over a large number of
iterations, with each one leading to closer approximations of an MSN.
The construction time can thus be adjusted with respect to the quality
of the graph.
\begin{itemize}
\item %
\emph{FANNG.} In the Fast ANN Graph \cite{harwood2016}, graph
construction terminates after a fixed number of trials, \textit{e.g.}
$50N$. The UpdateOutNeighbors routine adds an edge between $t$
and the nearest node in the search path, $p^*\in P$, and then 
prunes out-neighbors of $p^*$ based on ``occlusion'' rules derived
from the triangle inequality in order to limit out-degrees.
The empirical storage and search complexities are reported to
be on the order of $O(DN^{1-\epsilon})$.
\end{itemize}

\hi{Fixed Trial.}
In fixed trial construction, all trials are conducted from a special
designated source node, sometimes called the ``navigating'' node. This
node also serves as the source for all online queries.  Normally, the
index is constructed by conducting one trial to each node in a single
pass over $S$.
The construction complexity is generally about $O(DN^{1+\epsilon}\log
N^\epsilon)$ where the logarithmic term represents the cost of the
search trials.
\begin{itemize}
\item %
\emph{Navigating Spreading-Out Graph.}
The NSG index \cite{fu2019} starts from an approximate KNNG. For
UpdateOutNeighbors, it uses an edge selection strategy based on lune
membership, similar to \cite{dearholt1988}. To guarantee that all
targets are reachable from the navigating node, it overlays a spanning
tree to connect any unreachable targets.
\item %
\emph{Vamana.}
To speed up construction, Vamana \cite{subramanya2019} begins with a
random graph instead of an approximate KNNG, and instead of checking
lune membership, UpdateOutNeighbors uses a simple distance-based
threshold similar to FANNG \cite{harwood2016}.
%
\end{itemize}


\begin{table*}[t]
\centering
\begin{threeparttable}[b]
\caption{Representative search indexes.}
\label{tab:indexes}
\begin{tabular}{llllccccc}
\toprule
\multirow[b]{2}*{Structure}
  & \multirow[b]{2}*{Index\tnote{a}}
  & \multirow[b]{2}*{Partitioning}
  & \multirow[b]{2}*{Residence}
  & \multicolumn{3}{c}{Complexity\tnote{b}}
  & \multirow[b]{2}*{Update\tnote{c}}
  & \multirow[b]{2}*{\shortstack{Error\\Bound}} \\

  &
  &
  &
  & Constr.
  & Space
  & Query
  &
  & \\
\midrule
Table
  & E$^2$LSH \cite{datar2004}
  & Space
  & Mem.
  & \cm 
  & \ch 
  & \cm 
  & Y
  & \cmark \\

  & FALCONN \cite{andoni2015optimal}
  & Space
  & Mem.
  & \cm 
  & \ch 
  & \cm 
  & R
  & \cmark \\

  & *\texttt{SQ}
  & Discrete
  & Mem.
  & \cm 
  & \cl 
  & \cm 
  & Y 
  & \xmark \\

  & *\texttt{PQ}
  & Clustering
  & Mem.
  & \cm 
  & \cl 
  & \cm 
  & R 
  & \xmark \\

  & *\texttt{IVFSQ}
  & Clustering
  & Mem.
  & \cm 
  & \cl 
  & \cm 
  & R 
  & \xmark \\

  & *IVFADC \cite{jegou2011product}
  & Clustering
  & Mem.
  & \cm 
  & \cl 
  & \cm 
  & R 
  & \xmark \\

  & SPANN \cite{chen2021}
  & Clustering
  & Disk
  & \cm 
  & \cm 
  & \cm 
  & R
  & \xmark \\
Tree
  & FLANN \cite{muja2009}
  & Space
  & Mem.
  & \ch 
  & \ch 
  & \cl 
  & R 
  & \xmark \\

  & RPTree \cite{dasgupta2008,dasgupta2013}
  & Space
  & Mem.
  & \cl 
  & \ch 
  & \cl 
  & R 
  & \cmark \\

  & *ANNOY
  & Space
  & Mem.
  & \cl 
  & \ch 
  & \cl 
  & R 
  & \xmark \\
Graph
  & NN-Descent (KGraph) \cite{dong2011}
  & Proximity
  & Mem.
  & \cm 
  & \cm 
  & \cm 
  & N
  & \xmark \\

  & EFANNA
  & Proximity
  & Mem.
  & \ch 
  & \cm
  & \cl 
  & N 
  & \xmark \\

  & FANNG \cite{harwood2016}
  & Proximity
  & Mem.
  & \ch 
  & \cm
  & \cm 
  & N
  & \xmark \\

  & NSG \cite{fu2019}
  & Proximity
  & Mem.
  & \ch 
  & \cm
  & \cl 
  & N
  & \xmark \\

  & Vamana (DiskANN) \cite{subramanya2019}
  & Proximity
  & Disk
  & \cm 
  & \cm
  & \cl 
  & N
  & \xmark \\

  & *HNSW \cite{malkov2020}
  & Proximity
  & Mem.
  & \cl 
  & \cm
  & \cl 
  & Y 
  & \xmark \\
\bottomrule
\end{tabular}

\begin{tablenotes}
\item[a] %
An asterisk (*) indicates supported by more than two commercial
VDBMSs.
\item[b] %
Based on theoretical results reported by authors, empirical results
reported by authors, or our own cursory analysis when no results are
reported.
Key for complexity columns, with $0\leq\epsilon\leq 1$ and a natural
constant, $K$:
\parbox{\textwidth}{
  \begin{itemize}
  \item %
  \textit{Construction.}
  High=worse than $O(DN^{1+\epsilon})$,
  Med.=$O(DN^{1+\epsilon})$,
  Low=$O(DN\log N)$;
  \item %
  \textit{Size.}
  High=$O(DN\cdot K)$ or worse,
  Med.=$O((D+K)\cdot N)$,
  Low=$O(N\log D)$;
  \item %
  \textit{Query.}
  High=worse than $O(DN^\epsilon)$,
  Med.=$O(DN^\epsilon)$,
  Low=$O(D\log N)$.
  \end{itemize}
}
\item[c] %
Y=data independent updates; R=updates with rebalancing; N=no updates.
\end{tablenotes}
\end{threeparttable}
\end{table*}

\subsubsection{Small World Graphs}

\begin{figure}[t]
\centering
\includegraphics[width=0.30\textwidth]{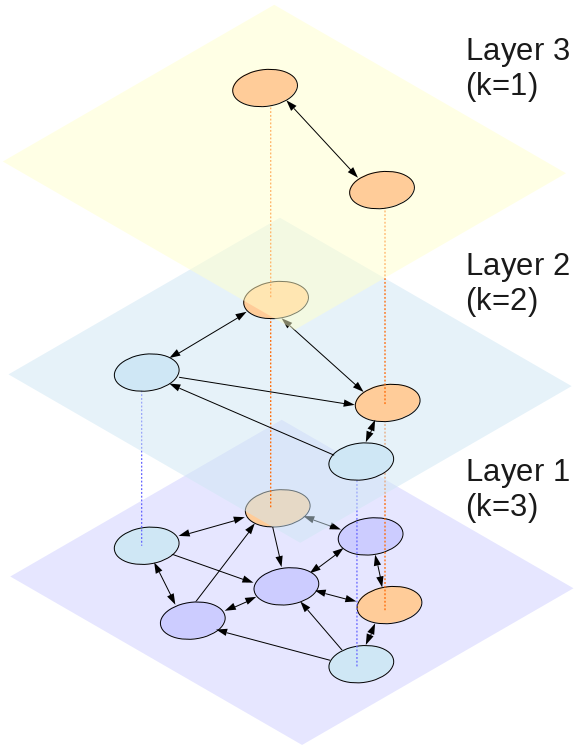}
\caption{Example HNSW index. Two nodes are randomly assigned to the
top layer (orange), two to the middle layer (blue), and three to the
bottom layer (purple). The out-degrees in each layer are bounded by
$k$ to modulate search complexity. Vertical edges allow the search
algorithm to traverse down the layers, while horizontal edges allow
traversal within a layer.}
\label{fig:hnsw}
\end{figure}

A graph is \emph{small-world} if the length of its characteristic path
grows in $O(\log N)$ \cite{watts1998}. A \emph{navigable} graph is one
where the length of the search path found by the best-first search
algorithm scales logarithmically with $N$ \cite{kleinberg2000}. A
graph that is both navigable and small-world (NSW) thus possesses a
search complexity that is likely to be logarithmic, even in the worst
case.
\begin{itemize}
\item %
\emph{NSW.}
An NSW graph can be constructed using a procedure which we call
\emph{one-shot refine} and detailed in \cite{malkov2014}.
Nodes are sequentially inserted into the graph, and when a
node is inserted, it is connected to its $k$ nearest neighbors already
in the graph.
\item %
\emph{HNSW.}
While NSW offers search paths that scale in $\log N$, the out-degrees
also tend to scale in the logarithm of $N$, leading to polylogarithmic
search complexity. In \cite{malkov2020}, a hierarchical NSW (HNSW)
graph is given which uses randomization in order to restore
logarithmic search.
During node insertion, the node is assigned to all layers below a
randomly selected maximum layer, chosen from an exponentially decaying
distribution so that the size of each layer grows logarithmically from
top to bottom. Within each layer, the node is connected to its
neighbors following the NSW procedure, but where the out-degrees are
bounded. Best-first search proceeds from the top-most layer.
Figure~\ref{fig:hnsw} shows an example.
\end{itemize}

\subsection{Discussion}

%
As seen from Table~\ref{tab:indexes}, HNSW offers many appealing
characteristics. It is easy to construct, has reasonable storage
requirements, can be updated, and supports fast queries. It therefore
comes as no surprise that it is supported by many commercial
VDBMSs. The storage cost may still be a concern for very large vector
collections, but there are ways to address this\footnote{For example,
Weaviate \cite{weaviate} allows constructing HNSW graphs over vectors
that have been compressed with PQ.}.

Even so, there are cases where other indexes may be more appropriate.
For batched queries or workloads where the queries belong to $S$,
KNNGs may be preferred, as once they are constructed, they can answer
these queries in $O(1)$ time. KGraph is easy to construct, but EFANNA
is more adaptable to any online queries.
For online workloads, the choice rests on several factors. If error
guarantees are important, then an LSH-based index or RPTree can be
considered. If memory is limited, then a disk-based index such as
SPANN or DiskANN may be appropriate. If the workload is write-heavy,
then table-based indexes may be preferred, as they generally can be
efficiently updated. Out of these, E$^2$LSH is data independent and
requires no rebalancing. For read-heavy workloads, tree or graph
indexes may be preferred, as they generally offer logarithmic
search complexity.

Aside from these indexes, there have also been efforts at mixing
structures in order to achieve better search performance.
For example, the index in \cite{azizi2023} and the Navigating Graph
and Tree (NGT) index \cite{ngt} use a tree to initially partition the
vectors and then use a graph index over each of the leaf nodes.

\section{Query Optimization and Execution}
\label{sec:opt}

There may be multiple ways to execute a given query. The goal of the
query optimizer is to select the optimal query plan, typically the
latency minimizing plan.

To achieve this goal, the first step is \emph{plan enumeration},
followed by \emph{plan selection} and then \emph{query execution}.
For predicated queries, vector indexes can not be easily combined with
attribute filters in the same plan, resulting in the development of
new \emph{hybrid} operators.


\subsection{Hybrid Operators}

Predicated queries can be executed by either applying
the predicate filter before vector search, known as ``pre-filtering'';
after the search, known as ``post-filtering''; or
during search, known as ``single-stage filtering''.

If the search is index-supported, then there needs to be a mechanism
to inform the index that certain vectors are filtered out.
For pre-filtering, \emph{block-first scan} works by ``blocking'' out
vectors in the index before the scan is conducted
\cite{wei2020,wang2021,gollapudi2023}. The scan itself proceeds as
normal but over the non-blocked vectors.
For single-stage filtering, \emph{visit-first scan} works by scanning
the index as normal, but meanwhile checking each visited vector
against the predicate conditions \cite{wu2022}.

\begin{figure}[t]
\centering
\includegraphics[width=0.3\textwidth]{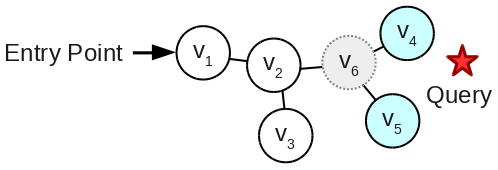}
\caption{Under best-first search, blocking out node $v_6$ makes the
blue cluster of nodes, comprising the true nearest neighbor,
unreachable from the entry point.}
\label{fig:bfs}
\end{figure}

\subsubsection{Block-First Scan}

Blocking can be done online at the time of a query, or if predicates
are known beforehand, it can be done offline.

\hi{Online Blocking.}
For online blocking, the aim is to perform the blocking as efficiently
as possible in order to minimize the impact on query latency.
In AnalyticDB-V \cite{wei2020} and Milvus \cite{milvus,wang2021}, a
technique using bitmasks is given. A bitmask is constructed using
traditional attribute filtering techniques. Then, during index scan, a
vector is quickly checked against the bitmask to determine whether it
is ``blocked''. 

\hi{Offline Blocking.}
For graph-based indexes, blocking can cause the graph to become
disconnected, as shown in Figure~\ref{fig:bfs}. In Filtered-DiskANN
\cite{gollapudi2023}, the aim is to prevent disconnections in the
first place by strategically adding edges based on the attribute
category of adjoining nodes. A similar preventative approach is used
in Qdrant \cite{qdrant} for HNSW and in \cite{wu2022}.

In Milvus \cite{milvus,wang2021}, $S$ is pre-partitioned along
attributes that are expected to be predicate targets. When a query
arrives, it can then be executed on the relevant partition using a
normal index scan.

\subsubsection{Visit-First Scan}


For low-selectivity predicates, visit-first scan can be faster than
online blocking because there is no need to block the vectors
beforehand.
But if the predicate is highly selective, then visit-first scan risks
frequent backtracking as the scan struggles to fill the result set.

One way to avoid backtracking is to infuse the scan operator with a
traversal mechanism that incorporates attribute information. In
\cite{gollapudi2023}, the filter condition is added to the best-first
search operator. In \cite{wu2022}, the distance function used for edge
traversal is augmented with an attribute related component so that the
scan favors nodes that are likely to pass the filter\footnote{See also
\url{http://arxiv.org/abs/2203.13601}.}.

\subsection{Plan Enumeration}

As vector search queries tend to consist of a small number of
operators, in many cases \emph{predefining} the plans is not only
feasible but also efficient, as it saves overhead of enumerating the
plans online.
But for systems that aim to support more complex queries,
the plans cannot be predetermined.
For extended VDBMSs based on relational systems, relational algebra
can be used to express these queries, allowing \emph{automatic}
enumeration.

\subsubsection{Predefined}

For predefined plans, the main consideration is which plan to specify
for which query. Some systems target specific workloads, thereby
focusing on single plans per query. Other systems predefine multiple
plans.

\hi{Single Plan.}
Single plans can be highly efficient as it removes the overhead of
plan selection in addition to enumeration, but can be a disadvantage
if the predefined plan is not suited to the particular workload.

A non-predicated query trivially has a single query plan when only one
type of search method is available.  For example in EuclidesDB
\cite{euclid}, each database instance is configured with one search
index which is used for every search query. This can also be true for
predicated queries.  For example in Weaviate, all predicated search
queries are executed by pre-filtering. Meanwhile in Vearch
\cite{vearch,li2018}, all predicated search queries are executed using
post-filtering.

\hi{Multiple Plans.}
For non-predicated queries, different indexes lead to multiple plans.
For example, AnalyticDB-V \cite{wei2020} supports brute force scan and
table-based index scan over \texttt{PQ} or VGPQ. This allows a {\KNN}
query to be executed using either of these methods.

Predicated queries can be answered either by pre-filtering,
post-filtering, or single-stage filtering. But different vector search
indexes, along with the presence or absence of an attribute index,
multiply the number of possible plans.

\subsubsection{Automatic}

For automatic enumeration, some VDBMSs based on relational systems
take advantage of the underlying relational optimizer to perform plan
enumeration as well as selection. The relational language is extended
to support distance functions and vector index scans.
For example, \texttt{pgvector} \cite{pgvec} and PASE \cite{yang2020}
both take advantage of PostgreSQL support for user extensions.

\subsection{Plan Selection}

To identify the optimal query plan, existing VDBMSs perform plan
selection either by using \emph{handcrafted rules} or by using a
\emph{cost model}.

\subsubsection{Rule Based}

\begin{figure*}[t]
\centering
\subfloat[][]{
  \label{fig:qdrant-rules}
  \includegraphics[width=0.40\textwidth]{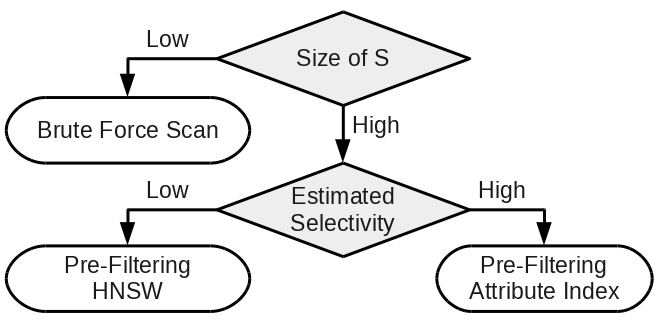}
}
\hspace{8pt}
\subfloat[][]{
  \label{fig:vespa-rules}
  \includegraphics[width=0.40\textwidth]{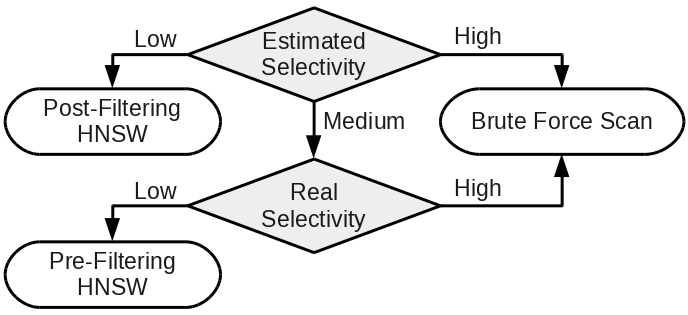}
}
\caption{Plan selection rules in (a) Qdrant and (b) Vespa.}
\label{fig:rules}
\end{figure*}

If the number of plans is small, then selection rules can be used to
decide which plan to execute. Figure~\ref{fig:rules} shows two
examples, used by Qdrant \cite{qdrant} (Figure~\ref{fig:qdrant-rules})
and Yahoo Vespa \cite{vespa} (Figure~\ref{fig:vespa-rules}).

Both Qdrant and Vespa perform plan selection based on the estimated
selectivity of the predicate.
In Qdrant, a predicated {\KNN} query can be executed by a brute force
scan, 
pre-filtering with an attribute index, 
or pre-filtering with HNSW.
Plan selection is based on two thresholds, one on the size of $S$ and
the other on the selectivity of the filter. If $S$ is small, then a
brute force scan is performed. On the other hand, if the selectivity
is high, meaning a small fraction of $S$ passes the filter, then the
attribute index is used, otherwise HNSW is used.
In Vespa, a predicated query can additionally be executed by using
HNSW followed by post-filtering
when the selectivity estimate is low. Moreover, if the estimate does
not immediately favor brute force scan or post-filtering, then Vespa
will apply the filter, and then choose between a brute force scan over
the filtered collection or a blocked HNSW scan based on the actual
selectivity.
Existing methods \cite{cormode2011} are used for estimating selectivity.

\subsubsection{Cost Based}

Plan selection can also be performed using a cost model, choosing the
plan with the least estimated cost.

In AnalyticDB-V \cite{wei2020} and Milvus \cite{milvus,wang2021}, a
linear cost model sums the component costs of the individual operators
to yield the cost of each plan. The basic operator cost depends on the
number of distance calculations as well as memory and disk retrievals
performed by the operator. For predicated queries, these numbers are
estimated from the selectivity of the predicate. But they also depend
on the desired query accuracy, which is exposed to the user as an
adjustable parameter. The effect of different accuracy levels on
operator cost is determined offline.

We point out several unaddressed challenges to cost estimation for
predicated queries. For pre-filtering with best-first search, the cost
of the scan can be hard to estimate due to uncertainty around the
amount of blocking. This applies more to tree or graph-based
indexes, as for table-based indexes, the cost of a scan is bounded
above by bucket size. Likewise for visit-first scan, the cost of the
scan depends on the rate of predicate failures, which is hard to know
beforehand.

On the other hand, post-filtering for a predicated {\KNN} query may
lead to a result set that contains fewer than $k$ items. In VDBMSs
that use post-filtering, this is often mitigated by retrieving $\alpha
k$ nearest vectors instead of just the $k$ nearest. But higher
$\alpha$ make search more expensive, and there is no clear way for
deciding on the optimal value which minimizes search cost while
guaranteeing $k$ results in the final result set.

\subsection{Query Execution}

Vector operators 
can take advantage of \emph{hardware acceleration} such as via
processor caches, SIMD instructions, and GPUs in order to reduce query
latency.
A VDBMS can also use \emph{distributed search} to reduce the computing
burden for a single machine, and to increase throughput, multiple
searches can be conducted in parallel over the distributed cluster.
For write-heavy workloads, a VDBMS can sacrifice consistency for write
throughput by using \emph{out-of-place updates}, which allow index
updates to be deferred until more suitable times.

\subsubsection{Hardware Acceleration}

Vector comparison requires reading full vectors into the processor.
This aspect of vectors along with large size complicates disk
retrieval, but this same locality makes them amenable to hardware
accelerated processing.

\begin{figure}[t]
\centering
\includegraphics[width=0.28\textwidth]{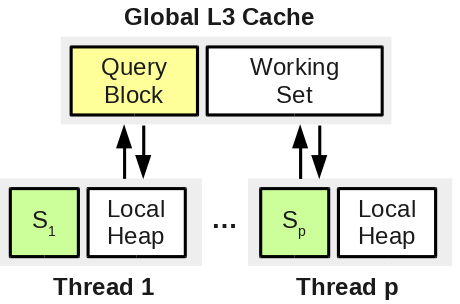}
\caption{Cache friendly query blocks used by Milvus
\cite{milvus,wang2021}.}
\label{fig:milvus-cache}
\end{figure}

\hi{CPU Cache.}
If data is not present in the processor cache, then it must be
retrieved from memory, stalling the processor.
As shown in Figure~\ref{fig:milvus-cache}, Milvus
\cite{milvus,wang2021} minimizes cache misses for batched queries by
partitioning the queries into query blocks, which are small enough to
fit into the CPU cache. The queries are answered a block at a time,
and multiple threads can be used to process the queries. As each
thread references the entire block when performing a search, the block
is safe from eviction under the common eviction policies.
%


\hi{Single Instruction Multiple Data (SIMD).}
The original ADC algorithm performs a series of table look-ups and
summations (see Example~\ref{ex:adc}). While SIMD instructions can
trivially parallelize the summations, look-ups require memory
retrievals (in the case of cache misses) and are more difficult to
speed up.
But in \cite{andre2017,andre2021}, the SIMD shuffle instruction is
cleverly exploited to parallelize these look-ups within a single SIMD
processor. This technique is implemented in Faiss \cite{faiss}.

\begin{figure}[t]
\centering
\includegraphics[width=0.28\textwidth]{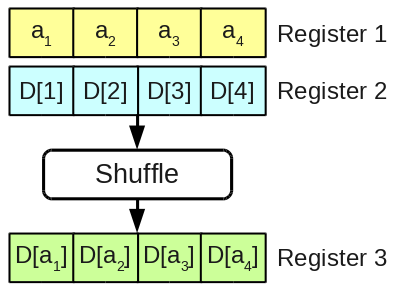}
\caption{Performing table look-up with SIMD.}
\label{fig:simd}
\end{figure}

The basic idea is illustrated in Figure~\ref{fig:simd}. The look-up
indices, plus the entire look-up table, are stored into the SIMD
registers. The shuffle operator is then used to rearrange the values
of the table register so that the $i$th entry contains the value at
the $i$th index, lining up the values for the subsequent
additions.

The table is aggressively compressed in order to fit it into a
register. First, centroids are represented using 4 bits instead of a
more customary 8 bits, yielding a total of $2^4=16$ centroids. Second,
the distances in the table are quantized into 8-bit integers. The
result is that there are only 16 8-bit integers in the look-up table,
fitting inside a typical 128-bit register. In \cite{andre2021},
some improvements are made to allow more values to be stored in the
register, namely variable-bit centroids and splitting up large tables
across multiple registers.

\hi{Graphical Processing Units (GPUs).}
A GPU consists of a large number of processing units in addition
to a large device memory. The threads within a processing unit are
grouped together in ``warps'', and each warp has access to a number of
32-bit registers shared across the threads. An architectural diagram
can be found in \cite{lindholm2008}.

In \cite{johnson2021}, an ADC search algorithm for GPUs is given, also
as a part of Faiss. Similar to the SIMD algorithm, the GPU algorithm
likewise tries to avoid memory retrievals, this time from GPU device
memory. It also achieves this by performing table look-ups within the
registers, taking advantage of a shuffle operator called ``warp
shuffle''.

If the running $k$ nearest neighbors are tracked in the registers,
then $k$ cannot be too large. Milvus \cite{milvus,wang2021} works
around this issue by conducting a {\KNN} over multiple rounds,
flushing intermediate results store in the registers into host memory
after each round.

\subsubsection{Distributed Search}

Many VDBMSs
\cite{vald,pine,milvus,wang2021,guo2022,qdrant,vespa,wei2020} take
advantage of distributed clusters in order to scale to larger datasets
or greater workloads.
%
%
Some that are offered as a cloud service take advantage of
disaggregated architectures in order to offer high elasticity.  We
point readers to \cite{wang2023disaggregated} for details about these
architectures.

To perform a distributed search, the vector collection is first
partitioned into shards. The collection can be partitioned into equal
shards, where the vectors in the shards are identically distributed,
or partitioned based on other characteristics, for example based on
index key for collections that are bucketed using a table-based index.
A local index can then be built for each shard, and
shards and their local indexes can also be replicated to provide fault
tolerance and to increase throughput, as multiple queries can be
executed simultaneously over the replicas.

Distributed vector search follows a scatter-gather pattern. The query
is first scattered to all the relevant shards, then the result set is
obtained by aggregating the results from each shard. For example for a
{\KNN} query, each shard produces a result set containing the $k$
nearest neighbors to the query that are members of the shard, and then
a central coordinator gathers and merges these results to produce the
final result set.


\subsubsection{Out-Of-Place Updates}

Updating an index in-place, even for fast update structures such as
HNSW, can disrupt search queries if the index cannot be used during
this time. These disruptions can be severe if updates take a long time
or during a lengthy rebuild.

\hi{Replicas.}
Some VDBMSs mitigate this issue by partitioning the vector collection
into shards and replicas, and then constructing a local index over
each replica \cite{vald,weaviate,wei2020}. In this way, if the index
on one replica is undergoing an update or rebuild, queries can be
handled by a different replica without any disruption. But the storage
(memory) requirements are multipled by the number of replicas, and
there may be extra overhead for search queries due to scatter-gather.

\hi{Log-Structured Merge (LSM) Tree.}
Another approach is to stream updates into a separate structure and
then reconciled against the index at a more convenient time.
An LSM tree \cite{oneil1996,luo2020} solves the problem that
read-friendly indexes cannot support fast writes, and write-friendly
update structures such as differential files and append-only logs
cannot support fast reads.

In Milvus \cite{milvus,wang2021} and Manu \cite{guo2022}, a vector
search index is constructed over each segment of the LSM tree in order
to support fast reads. A new index is created whenever a segment
becomes full or is merged.

\hi{Other Techniques.}
In Vald \cite{vald}, updates are streamed into a simple queue, and
then applied in bulk to the local index once the queue is full or
based on other conditions. Likewise in AnalyticDB-V \cite{wei2020},
updates are kept in memory and periodically merged with older
records that are kept on disk.

%


\section{Current Systems}
\label{sec:sys}




\begin{table*}[t]
\centering
\begin{threeparttable}[b]
\caption{A list of current VDBMSs. Dates are estimated from paper
publication, earliest Github release, blog post, or other
indications. Dates indicate year when vector search capability first
appeared in the product.}
\label{tab:vdbms}
\begin{tabular}{lllaaacccaaac}
\toprule
\multirow{2}*{Name}
  & \multirow{2}*{Type}
  & \multirow{2}*{\shortstack{Sub-\\Type}}
  & \multicolumn{3}{c}{Vector Query}
  & \multicolumn{3}{c}{Query Variant}
  & \multicolumn{3}{c}{Vector Index}
  & \\

  &
  & 
  & Ex.
  & Ap.
  & Rng.
  & Pr.
  & Mul.
  & Bat.
  & Tab.
  & Tr.
  & Gr.
  & Opt. \\
\midrule
EuclidesDB (2018) \cite{euclid}
  & Nat.
  & Vec.
  & \cmark
  & \cmark
  & \xmark
  & \xmark
  & \xmark
  & \xmark
  & \cmark
  & \cmark
  & \cmark
  & \xmark \\
Vearch (2018) \cite{vearch,li2018}
  & Nat.
  & Vec.
  & \xmark
  & \cmark
  & \xmark
  & \cmark
  & \xmark
  & \xmark
  & \cmark
  & \xmark
  & \xmark
  & \xmark \\
Pinecone (2019) \cite{pine}
  & Nat.
  & Vec.
  & \xmark
  & \cmark
  & \xmark
  & \cmark
  & \cmark
  & \xmark
  & \multicolumn{3}{a}{Proprietary}
  & U \\
Vald (2020) \cite{vald}
  & Nat.
  & Vec.
  & \cmark
  & \cmark
  & \cmark
  & \xmark
  & \xmark
  & \cmark
  & \xmark
  & \xmark
  & \cmark
  & \xmark \\
Chroma (2022) \cite{chroma}
  & Nat.
  & Vec.
  & \xmark
  & \cmark
  & \xmark
  & \cmark
  & \xmark
  & \xmark
  & \xmark
  & \xmark
  & \cmark
  & U \\
Weaviate (2019) \cite{weaviate}
  & Nat.
  & Mix
  & \xmark
  & \cmark
  & \cmark
  & \cmark
  & \cmark
  & \xmark
  & \xmark
  & \xmark
  & \cmark
  & \xmark \\
Milvus (2021) \cite{milvus,wang2021}
  & Nat.
  & Mix
  & \cmark
  & \cmark
  & \cmark
  & \cmark
  & \cmark
  & \cmark
  & \cmark
  & \xmark
  & \cmark
  & \cmark \\
NucliaDB (2021) \cite{nuclia}
  & Nat.
  & Mix
  & \xmark
  & \cmark
  & \cmark
  & \cmark
  & U
  & \xmark
  & \xmark
  & \xmark
  & \cmark
  & \xmark \\
Qdrant (2021) \cite{qdrant}
  & Nat.
  & Mix
  & \cmark
  & \cmark
  & \cmark
  & \cmark
  & \cmark
  & \cmark
  & \xmark
  & \xmark
  & \cmark
  & \cmark \\
Manu (2022) \cite{guo2022}
  & Nat.
  & Mix
  & \cmark
  & \cmark
  & \cmark
  & \cmark
  & \cmark
  & \cmark
  & \cmark
  & \xmark
  & \cmark
  & \cmark \\
Marqo (2022) \cite{marqo}
  & Nat.
  & Mix
  & \xmark
  & \cmark
  & \xmark
  & \cmark
  & \cmark
  & \cmark
  & \xmark
  & \xmark
  & \cmark
  & U \\
\midrule
Vespa (2020) \cite{vespa}
  & Ext.
  & NoSQL
  & \cmark
  & \cmark
  & \cmark
  & \cmark
  & \cmark
  & \xmark
  & \xmark
  & \xmark
  & \cmark
  & \cmark \\
Cosmos DB (2023)
  & Ext.
  & NoSQL
  & \xmark
  & \cmark
  & \xmark
  & \xmark
  & \xmark
  & \xmark
  & \cmark
  & \xmark
  & \xmark
  & \xmark \\
MongoDB (2023)
  & Ext.
  & NoSQL
  & \xmark
  & \cmark
  & \xmark
  & \cmark
  & \xmark
  & \xmark
  & \xmark
  & \xmark
  & \cmark
  & \xmark \\
Neo4j (2023)
  & Ext.
  & NoSQL
  & \xmark
  & \cmark
  & \xmark
  & \xmark
  & \xmark
  & \xmark
  & \xmark
  & \xmark
  & \cmark
  & \xmark \\
Redis (2023)
  & Ext.
  & NoSQL
  & \cmark
  & \cmark
  & \cmark
  & \cmark
  & \xmark
  & \cmark
  & \xmark
  & \xmark
  & \cmark
  & \xmark \\
AnalyticDB-V (2020) \cite{wei2020}
  & Ext.
  & Rel.
  & \cmark
  & \cmark
  & \cmark
  & \cmark
  & \xmark
  & \xmark
  & \cmark
  & \xmark
  & \cmark
  & \cmark \\
PASE+PG (2020) \cite{yang2020}
  & Ext.
  & Rel.
  & \cmark
  & \cmark
  & \cmark
  & \cmark
  & \xmark
  & \xmark
  & \cmark
  & \xmark
  & \cmark
  & \cmark \\
pgvector+PG (2021) \cite{pgvec}
  & Ext.
  & Rel.
  & \cmark
  & \cmark
  & \cmark
  & \cmark
  & \cmark
  & \xmark
  & \cmark
  & \xmark
  & \cmark
  & \cmark \\
SingleStoreDB (2022) \cite{single,prout2022}
  & Ext.
  & Rel.
  & \cmark
  & \xmark
  & \cmark
  & \cmark
  & \xmark
  & \xmark
  & \xmark
  & \xmark
  & \xmark
  & \cmark \\
ClickHouse (2023) \cite{click}
  & Ext.
  & Rel.
  & \cmark
  & \cmark
  & \cmark
  & \cmark
  & \xmark
  & \xmark
  & \xmark
  & \cmark
  & \cmark
  & \cmark \\
MyScale (2023) \cite{my}
  & Ext.
  & Rel.
  & \cmark
  & \cmark
  & \cmark
  & \cmark
  & \xmark
  & \xmark
  & \cmark
  & \xmark
  & P
  & \cmark \\
\bottomrule
\end{tabular}
\begin{tablenotes}[para]
\footnotesize Abbreviations: 
Ex.=exact {\KNN};
Ap.=ANN;
Rng.=range;
Pr.=predicated;
Mul.=multi-vector;
Bat.=batched;
Tab.=table;
Tr.=tree;
Gr.=graph;
Opt.=query optimizer;
Nat.=native;
Ext.=extended;
Vec.=mostly vector;
Rel.=relational;
U=unknown;
P=proprietary.
\end{tablenotes}
\end{threeparttable}
\end{table*}

The variety of data management techniques has led to an equally
diverse landscape of commercial VDBMSs. They can be broadly
categorized as \emph{native} systems, which are designed specifically
for vector data management, and \emph{extended} systems, which add
vector capabilities on top of an existing system.
Table~\ref{tab:vdbms} lists several of these systems.

\subsection{Native}

Native systems are characterized by small query APIs, simple
processing flows composed of a low number of components, and basic
storage models, making them highly specialized at vector data
management.

These systems can be further divided into two subcategories, those
that target \emph{mostly vector} workloads
\cite{euclid,vearch,pine,vald,chroma}, where the vast majority of
queries access the vector collection, and those that target
\emph{mostly mixed} workloads
\cite{weaviate,milvus,nuclia,qdrant,guo2022,marqo}, where queries are
also expected to access non-vector collections. Mostly mixed workloads
may consist of traditional attribute queries or textual keyword
queries, along with typical predicated and non-predicated vector
queries.

\subsubsection{Mostly Vector}

Mostly-vector systems are designed to support fast search queries over
large vector collections. Several systems, such as EuclidesDB
\cite{euclid} and Vald \cite{vald}, focus exclusively on
non-predicated queries. Others, such as Vearch \cite{vearch}, Pinecone
\cite{pine}, and Chroma \cite{chroma} offer predicated queries but
supported by a single predefined plan.
These systems often support only a single search index, typically
graph-based.
Hence, they have no need for a query parser, rewriter, or optimizer,
and can omit these components in order to reduce processing
overhead. They also typically do not support exact search, as all
queries are handled by the index.

\begin{figure}[t]
\includegraphics[width=0.45\textwidth]{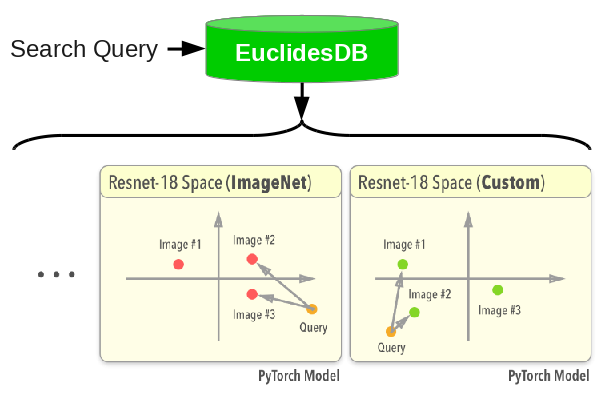}
\caption{EuclidesDB \cite{euclid} supports multiple vector spaces
so that query results can be more semantically accurate.}
\label{fig:euclides}
\end{figure}

We point out some examples.
\begin{itemize}
\item %
\emph{EuclidesDB.}
EuclidesDB \cite{euclid}, shown in Figure~\ref{fig:euclides}, is a
system for managing embedding models, aimed at allowing users to
experiment over various models and similarity scores in order to find
the most semantically meaningful setting. Users bring their own
embedding models and interact with the system by querying and
manipulating non-vector entities.
\item %
\emph{Vald.}
Vald \cite{vald} is a serverless VDBMS aimed at providing scalable
non-predicated vector search. Instead of a single centralized system,
Vald spans a Kubernetes\footnote{\url{http://kubernetes.io}} cluster
consisting of multiple ``agents''. The vector collection is sharded
and replicated across the agents to increase throughput and decrease
latency via scatter-gather. 
An individual Vald agent contains an NGT graph index built over its
local shard. Vald also supports out-of-place updates on each local
replica using a simple update queue.
\item %
\emph{Vearch.}
Vearch \cite{vearch,li2018} is targeted at image-based search for
e-commerce, allowing it to be designed specifically toward this
aim. Predicated queries are supported, but they are all
executed using post-filtering, considered to be sufficient for the
application. Vearch also has no need to support exact queries.
Vearch adopts a disaggregated architecture with dedicated search nodes,
allowing it to scale read and write capabilities independently.
\end{itemize}
Pinecone \cite{pine} offers a scalable distributed system similar to
Vald, but as it is offered as a managed cloud-based service, it can be
a more user-friendly option. Chroma \cite{chroma} is a centralized
system similar to EuclidesDB, meant for a single machine.

\subsubsection{Mostly Mixed}

Mostly-mixed systems aim to support more sophisticated queries
compared to mostly-vector systems. In general, they support a greater
variety of basic queries, including more support for exact and range
queries, in addition to more query variants. These systems likewise
have no need for query rewriters or parsers, but some,
such as Milvus \cite{milvus,wang2021}, Qdrant \cite{qdrant},
and Manu \cite{guo2022}, make use of query optimization. Other
systems, including Weaviate \cite{weaviate}, NucliaDB \cite{nuclia},
and Marqo \cite{marqo}, also have more sophisticated data and storage
models in order to deal with predicated queries and attribute-only
queries that go beyond simple retrieval.

\begin{itemize}
\item %
\emph{Milvus and Manu.} Milvus \cite{milvus,wang2021} is aimed at
comprehensive support for vector search queries, and Manu
\cite{guo2022} adds additional features on top of Milvus. All three
basic query types are supported, in addition to the three query
variants. Multiple search indexes are also supported, and predicated
queries are handled by a cost-based optimizer.
\item %
\emph{Qdrant.} Qdrant \cite{qdrant} likewise supports a large variety
of search queries. For predicated queries, it uses a rule-based
optimizer coupled with a custom HNSW index designed for block-first
scan.
\item %
\emph{NucliaDB and Marqo.}
NucliaDB \cite{nuclia} and Marqo \cite{marqo} are targeted at document
search and retrieval and use vector search to provide semantic
retrieval. A key feature is the support for combining keywords and
vectors through multi-vector search.
Non-vector keyword queries are processed using text specific
techniques, resulting in sparse term-frequency vectors. These vectors
are then combined with dense feature vectors through an aggregate
score in order to conduct multi-vector search.
\item %
\emph{Weaviate.}
Weaviate targets document search and retrieval over a graph model.
This allows Weaviate to answer non-vector queries, such as retrieving
all books written by a certain author, in addition to similarity
queries via vector search. Users interact with Weaviate by issuing
queries written in GraphQL\footnote{http://graphql.org}.
\end{itemize}

\subsection{Extended}

Extended systems inherit all the capabilities of an underlying data
management system and are necessarily more complex compared to native
systems. But these systems are also more capable. Nearly all the
extended systems listed in Table~\ref{tab:vdbms} support all three
basic query types and multiple indexes, and all of them support query
optimization.
These systems likewise divide into two subcategories, those where the
underlying system is NoSQL and those where it is relational.

\subsubsection{NoSQL}


Many of the hallmarks of a NoSQL system \cite{davoudian2018}, such as
schemaless storage, distributed architecture, and eventual
consistency, are present in native systems, making a NoSQL extended
VDBMS much like a native system.
\begin{itemize}
\item %
\emph{Vespa.}
Vespa \cite{vespa} is a scalable distributed NoSQL system designed for
large scale data processing workloads. Unlike native systems, Vespa
aims at more general data processing tasks, hence it is equipped with
a flexible SQL-like query language. But like native systems, the
storage model is simple, provided by an in-house document store. Vespa
uses a rule-based query optimizer.
\item %
\emph{Cassandra.}
Cassandra \cite{lakshman2010} is a popular distributed NoSQL system
based on a wide column store. Vector search capability will be
available in version 5.0\footnote{See Cassandra Enhancement Proposal
30 (CEP-30).} by integrating HNSW into the storage layer, implementing
scatter-gather across replicas, and extending the Cassandra query
language with vector search operators.
\item %
\emph{Databricks.}
The Spark-based Databricks\footnote{\url{http://databricks.com}}
platform is expected to support vector search, including predicated
queries, in an upcoming version.
\end{itemize}

Aside from Vespa, several other document based NoSQL databases have now
been extended to support vector search, including
MongoDB\footnote{\url{http://mongodb.com}}, Cosmos
DB\footnote{\url{http://cosmos.azure.com}}, and
Redis\footnote{\url{http://redis.io}}, when coupled with the Redis
Stack search extension.
%
%
Vector search capability has also been extended to NoSQL systems other
than document stores. For example,
Neo4j\footnote{\url{http://neo4j.com}} is a property graph database
with experimental vector search capability. So far, only {\ANN}
queries are supported, using HNSW.

\subsubsection{Relational}


For extended relational systems, their features mostly come from
the inherent capabilities of relational systems. For example, SQL
already is sufficient for expressing $(c,k)$-search queries, and
these queries can already be answered by most relational engines upon
adding a user-defined similarity function. This is perhaps exemplified
by SingleStore \cite{single,prout2022}, which offers exact {\KNN} and
range search through its native relational engine, without any vector
search indexes. Subsequently, extended relational systems focus more
on tightly integrating vector search capability alongside existing
components.
\begin{itemize}
\item %
\emph{SingleStore.}
SingleStore \cite{single,prout2022} is a cloud-native hybrid
transactional and analytical (HTAP) database \cite{li2022} based on a
distributed row store for fast writes alongside column store replicas
for fast reads. Vector search is handled by the native relational
engine, extended with functions for calculating dot product and
Euclidean distance.
\item %
\emph{PASE.}
PASE \cite{yang2020} extends PostgreSQL with a flat quantization index
and HNSW index in order to support vector search.
\item %
\emph{\texttt{pgvector}.}
The \texttt{pgvector} \cite{pgvec} extension for PostgreSQL brings a
vector data type, functions over this data type, and flat and HNSW
index access methods into PostgreSQL. Vector queries are issued using
SQL. If an index is created over a vector column, then queries are
answered using the index, yielding approximate results. Otherwise,
exact brute force scan is used. Plan selection is performed by the
existing PostgreSQL query optimizer using the generic cost estimator,
or by calling index-specific cost estimators to refine the estimate,
if needed. Other features such as replication, fault tolerance,
access controls, concurrency control, and so on, are provided by
PostgreSQL.
\item %
\emph{AnalyticDB-V.}
AnalyticDB-V \cite{wei2020} adds vector search capability on top of
AnalyticDB \cite{zhan2019}.  This is achieved primarily by introducing
vector indexes, VGPQ and HNSW, and by augmenting the cost-based
optimizer.
\item %
\emph{ClickHouse and MyScale.}
ClickHouse \cite{click} is a columnar database aimed at fast analytics
coupled with an asynchronous merge mechanism for fast ingestion. It
supports vector queries using ANNOY and HNSW. Likewise, MyScale
\cite{my} is a cloud service using ClickHouse as the backend. It adds
table-based search indexes including flat indexes and IVFADC, and a
proprietary search index called ``multi-scale tree graph'' (MSTG)
which is shown to outperform both IVFADC and HNSW.
\end{itemize}


\subsection{Libraries and Other Systems}

Vector search engines and libraries make up a third category in the
broader ecosystem. Typically, these are embedded into applications
that require vector search, but they lack the capabilities of a full
VDBMS.

\hi{Search Engines.}
Apache Lucene \cite{lucene} is a pluggable search engine that provides
sophisticated search capability for embedded applications. Latest
versions offer vector search, supported by HNSW. While Lucene itself
lacks higher level features such as multi-tenancy, distributed search,
and administrative features, many of these are provided by search
platforms built on top of Lucene, including Elasticsearch
\cite{elastic}, OpenSearch \cite{open}, and Solr \cite{solr}. These
capabilities could make Lucene an attractive alternative to
mostly-vector native VDBMSs\footnote{For a discussion, see
\url{http://arxiv.org/abs/2308.14963}} as it offers similar features
and can readily integrate with existing infrastructure.

\hi{Libraries.}
There are also libraries that directly implement specific indexes. For
example, KGraph is an implementation of the NN-Descent KNNG.
The Microsoft Space Partition Tree and Graph (SPTAG) library
\cite{sptag} combines several techniques, including SPANN and NGT,
into one configurable index.
Libraries are also available for LSH, including
E2LSH\footnote{\url{http://www.mit.edu/~andoni/E2LSH_gpl.tar.gz}} and
FALCONN\footnote{\url{http://github.com/falconn-lib/falconn}}.
Likewise, Meta Faiss \cite{faiss} offers a selection of indexes,
including HNSW, an LSH family for Hamming distance, and
quantization-based indexes.

\hi{Other Systems.}
Other systems are aimed at other parts of the broader
pipeline. Featureform\footnote{\url{http://featureform.com}} is a
middleware aimed at dataset management. Similar to relational ETL
tools, it offers a means of organizing the workflows that transform
raw data sources into curated datasets to be consumed by downstream
applications. Featureform exposes some of these downstream functions
through its API. For example, it offers a vector search endpoint that
executes a {\KNN} query over a configured provider, such as Pinecone.
On the other hand, Activeloop\footnote{\url{http://activeloop.ai}}
Deep Lake \cite{hambardzumyan2023} offers tensor operations directly
over a tensor warehouse, making it capable of performing vector search
inside the warehouse.

Aside from vector search, vector visualization can also be a useful
tool for data analysts. Nomic Atlas\footnote{\url{http://nomic.ai}}
offers a platform for visualizing high-dimensional vector spaces by
projecting them onto two-dimensional plots using a proprietary
dimensionality reduction technique.

\subsection{Discussion}

\begin{figure}[t]
\centering
\includegraphics[width=0.45\textwidth]{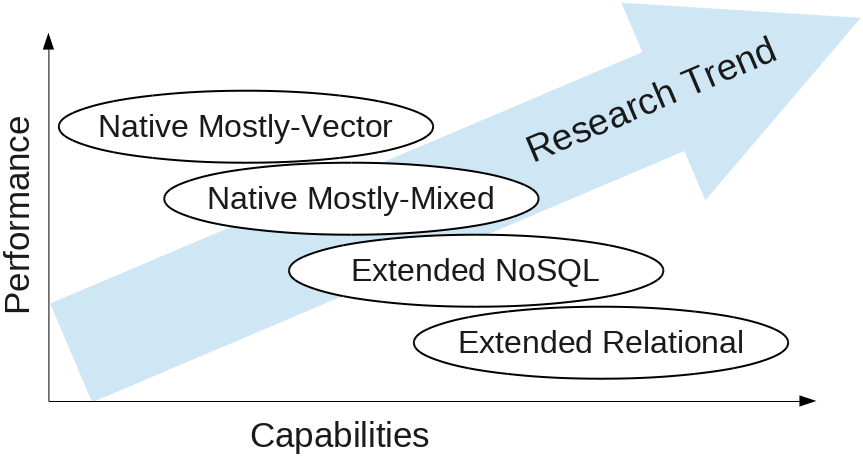}
\caption{High-level characteristics of VDBMSs.}
\label{fig:summary}
\end{figure}

The designs of these databases cover a spectrum of characteristics
involving query processing and vector storage, manifesting in a range
of performance and capabilities, as shown in Figure~\ref{fig:summary}.

We offer a few broad remarks. Native mostly-vector systems broadly
offer high performance but are targeted at specific workloads,
sometimes even specific queries, and thus have relatively limited
capability. Meanwhile, native mostly-mixed systems offer more
capabilities, notably predicated queries, and some such as Milvus
\cite{milvus,wang2021}, Qdrant \cite{qdrant}, and Manu \cite{guo2022}
also perform query optimization. These, along with extended NoSQL
systems, achieve a comfortable balance between high performance and
search capabilities. On the other hand, extended relational systems
offer the most capabilities but possibly less performance. But, as has
been mentioned
elsewhere\footnote{\url{https://arxiv.org/abs/2308.14963}}, relational
systems are already major components of industrial data
infrastructures, and being able to conduct vector search without
introducing new systems into the infrastructure is a compelling
advantage. The ranking shown in Figure~\ref{fig:summary} is
consistent with empirical observations \cite{aumuller2020}.

Thus, we imagine that future work will target systems that can offer
both high performance in addition to offering unified data management
capabilities, represented by the arrow in Figure~\ref{fig:summary}.

\section{Benchmarks}
\label{sec:bench}

Comprehensive cross-disciplinary comparisons of vector search
algorithms and systems are surprisingly scarce, attributed to the wide
range of fields from which search algorithms arise \cite{li2020}. We
point out two notable attempts at benchmarking these algorithms and
systems.

In \cite{li2020}, a large number of {\ANN} algorithms are uniformally
implemented and evaluated across a range of experimental
conditions. These algorithms include LSH, L2H, methods based on
quantization, tree-based techniques,
and graph-based techniques. 
The experiments are conducted over 18 datasets, ranging from a
few thousand vectors to 10 million vectors, and with dimensions
ranging from 100 to 4,096. The feature vectors are derived from
real-world image, text, video, and audio collections, as well as
synthetically generated. Algorithms are measured on query latency as
well as the quality of the result sets based on precision, recall, and
two other derivative measures.

In \cite{aumuller2020}, the evaluation is extended to include full
VDBMSs in addition to isolated algorithms. Whereas \cite{li2020} aims
to avoid the effects of different implementations, here these
differences are kept in order to more accurately reflect real-world
conditions. The standardized evaluation platform and the latest
benchmark results are available
online\footnote{\url{http://ann-benchmarks.com}}, with results for
several VDBMSs.
The datasets, however, are smaller in both size and dimensionality
than the ones used in \cite{li2020}.



\section{Challenges and Open Problems}
\label{sec:disc}

While much progress has been made on vector data management, some
challenges remain unaddressed.

\hi{Similarity Score Selection.}
The semantic quality of different similarity scores is still
challenging to understand, and there is no rigorous guidance on how to
select which score for which scenarios. Systems like EuclidesDB
\cite{euclid} can be used to experimentally determine the best scores
and embedding models, but outside of \cite{wang2023}, this problem
remains unexplored.

\hi{Operator Design.}
Designing efficient and effective hybrid operators remains
challenging. For graph indexes, block-first scan can lead to
disconnected components that either need to be repaired or that
require new search algorithms for handling this situation.
Existing offline blocking techniques
\cite{gollapudi2023,milvus,wang2021,qdrant} are limited to small
number of attribute categories.
For visit-first scan, estimating the cost of the scan is challenging
due to unpredictable backtracking, complicating plan selection.

\hi{Incremental Search.}
Some applications, such as e-commerce and recommender platforms, make
use of incremental {\KNN} search, where $k$ is effectively very large
but is retrieved in small increments so that the results appear to be
seamlessly delivered to the user. While techniques exist for
this type of search \cite{fisher2012}, it is unclear how to
support this search inside vector indexes.

\hi{Multi-Vector Search.}
Multi-vector search is also important for certain applications such as
face recognition. Existing techniques tend to use aggregate scores,
but this can be inefficient as it multiplies the amount of distance
calculations. Generic multi-attribute top-$k$ techniques also are hard
to adapt to vector indexes \cite{wang2021}, and there are no works on
MQMF queries.

%

\hi{Security and Privacy.} As vector search becomes increasingly
mission-critical, data security and user privacy become more
important, especially for VDBMSs that offer managed cloud services.
There is thus a need for new techniques that can support private and
secure high-dimensional vector search \cite{xue2018}.

\section{Conclusion}
\label{sec:conc}

In this paper, we surveyed vector database management systems aimed at
fast and accurate vector search, developed in response to recent
popularity of dense retrieval for applications such as LLMs and
e-commerce. We reviewed considerations for query processing, including
similarity scores, query types, and basic operators. We also reviewed
the design, search, and maintenance considerations regarding vector
search indexes. We described several techniques for query optimization
and execution, including plan enumeration, plan selection, operators
for predicated or ``hybrid'' queries, and hardware acceleration.
Finally, we discussed several commercial systems and the main
benchmarks for supporting experimental comparisons.


\bibliographystyle{spmpsci}
\bibliography{main}

\end{document}